\documentclass[graybox]{svmult}

\usepackage{type1cm}        
\usepackage{makeidx}         
\usepackage{graphicx}        
\usepackage{multicol}        
\usepackage[bottom]{footmisc}

\usepackage{newtxtext}       %
\usepackage{newtxmath}       
\newcommand{\ep}{\varepsilon}
\newcommand{\Li}[2]{{\mbox{Li}}_{#1}\left(#2\right)}
\newcommand{\Snp}[2]{{\mbox{S}}_{#1\!}\left(#2\right)}
\makeindex             

\begin{document}
	
	\title*{Hypergeometric Functions and Feynman Diagrams}
	\author{Mikhail Kalmykov, Vladimir Bytev, 
		   Bernd Kniehl,  Sven-Olaf~Moch, Bennie Ward, and Scott Yost}
	\institute{Mikhail Kalmykov  \at 
		JINR, Dubna, Russia, 
	\email{kalmykov.mikhail@gmail.com}
	 \and 
		Vladimir Bytev 
         \at BLTP JINR, 141980 Dubna, Moscow region, Russia 
	 \email{bytev@protonmail.com}
      \and 
         Bernd A. Kniehl, 
       \at
       II. Institut fuer Theoretische Physik, 
       Universitaet Hamburg,
       Luruper Chaussee 149, 
       22761 Hamburg, Germany
       \email{kniehl@desy.de}
       \and 
        Sven-Olaf~Moch
        \at 
         II. Institut fuer Theoretische Physik, 
        Universitaet Hamburg,
        Luruper Chaussee 149, 
        22761 Hamburg, Germany
        \email{sven-olaf.moch@desy.de}
       \and  
         B.F.L.~Ward
          \at 
          Department of Physics, Baylor University,
          One Bear Place, \# 97316, Waco, TX 76798-7316, USA
           \email{BFL{\underline\ }Ward@baylor.edu} 
        \and  
            S.A.~Yost 
           \at 
           Department of Physics, The Citadel, 171 Moultrie St.,
           Charleston, SC 29409, USA
   	\email{scott.yost@citadel.edu}
   \and 
 }
	%
	%
	\maketitle

\abstract{The relationship between Feynman diagrams and hypergeometric functions
is discussed. Special attention is devoted to existing techniques for the 
construction of the $\ep$-expansion. As an example, we present a detailed 
discussion of the construction of the $\ep$-expansion of the Appell function 
$F_3$ around rational values of parameters via an iterative solution of 
differential equations.  
As a by-product, we have found that the one-loop massless pentagon diagram 
in dimension $d=3-2\epsilon$ is not expressible in terms of multiple polylogarithms. 
Another interesting example is the Puiseux-type 
solution involving a differential operator generated by a hypergeometric 
function of three variables. 
The holonomic properties of the $F_N$ hypergeometric 
functions are briefly discussed.} 

\section{Introduction}
\label{Introduction}
Recent interest in the analytical properties of 
Feynman diagrams has been motivated by processes at the LHC. 
The required precision demands the evaluation of a
huge number of diagrams having many scales to a high order, 
so that a new branch of mathematics emerges, which we may call 
{\it the Mathematical Structure of Feynman Diagrams}
~\cite{hwa,pham}, which includes elements of algebraic geometry, algebraic 
topology, the analytical theory of differential equations, multiple 
hypergeometric functions, 
elements of number theory, modular functions and elliptic curves, 
multidimensional residues, and graph theory.     
This mathematical structure has been extensively developed, studied and 
applied. For a more detailed discussion of the oldest results and their relation 
to modern techniques, see
Refs.\ \cite{golubeva}\ and \cite{Kalmykov:2008}). 
One of these approaches is based on the treatment of Feynman 
diagrams in terms of multiple hypergeometric 
functions~\cite{kershaw}. For example, in the series of 
papers~\cite{kreimer1,kreimer2,kreimer3}, the one-loop diagrams 
have been associated with the $R$-function (a particular case 
of the $F_D$-function~\cite{bateman,slater,srivastava}).

\subsection{Mellin-Barnes representation, 
	asymptotic expansion, NDIM}
\label{Mellin-Barnes representation}

A universal technique based on the Mellin-Barnes representation of Feynman 
diagrams has been applied to one-loop diagrams in 
Ref.\ \cite{boos-davydychev,davydychev:1991}
and to two-loop propagator diagrams in 
Ref.\ \cite{broadhurst,berends,bauberger,davydychev-grozin,weinzierl:2003}.\footnote{Several 
	programs are available for the automatic generation of the  
	Mellin-Barnes representation of Feynman 
	diagrams~\cite{ambre,smirnov-smirnov,prausa}.}
The multiple Mellin-Barnes representation for a Feynman diagram 
in covariant gauge can be written in the form
\begin{eqnarray}
	\Phi({\bf A},\vec{B};{\bf C}, \vec{D};\vec{z}) 
	& = & 
	\int_{-i \infty}^{+i \infty}
	\phi(\vec{t}) 
	d\vec{t} 
	= 
	\int_{-i \infty}^{+i \infty}
	\prod_{a,b,c,r}
	\frac{\Gamma(\sum_{i=1}^m A_{ai}t_i \!+\! B_a)}{\Gamma(\sum_{j=1}^r 
	C_{bj}t_j \!+\! D_{b})}
	dt_c z_k^{\sum_l \alpha_{kl} t_l}
	\;,
	\nonumber \\ 
	\label{MB}
\end{eqnarray}
where $z_k$ are ratios of Mandelstam variables and 
$A,B,C,D$ are matrices and vectors depending linearly on the 
dimension of space-time $n$ and powers of the propagators. 
Closing the contour of integration on the right
(on the left), this integral can be presented  
around zero values of $\vec{z}$ in the form
\begin{eqnarray}
	&& 
		\Phi({\bf A},\vec{B};{\bf C}, \vec{D};\vec{z}) 
	= 
	\sum_{\vec{\alpha}} f_{\vec{\alpha}}
	H({\bf A},\vec{B};{\bf C}, \vec{D};\vec{z}) 
	\vec{z}^{\;\vec{\alpha}} 
	\;, 
	\label{representation}
\end{eqnarray}	
where the coefficients 
$
f_{\vec{\alpha}}
$
are ratios of $\Gamma$-functions and the functions $H$ are 
Horn-type hypergeometric functions~\cite{horn} 
(see Section~\ref{Horn-Functions} for details).
The analytic continuation of the hypergeometric functions 
$H(\vec{z})$ into another region of the variables $\vec{z}$
can be constructed via the integral representation 
(when available)~\cite{wu,mano}, 
$H(\vec{z}) \to H(1-\vec{z})$. 
However, for more complicated cases of Horn-type hypergeometric
functions, this type of analytic continuation is still under 
construction~\cite{friot1,friot2}.

A major set of mathematical results (see, for example,~\cite{bezrodnykh1,bezrodnykh2,bezrodnykh3}) is devoted to the 
construction of the analytic continuation of a series around ${z_j}=0$ 
to a series of the form
$\frac{z_A}{z_B}$: $H(\vec{z}) \to H(\frac{z_A}{z_B})$, 
where the main physical application is the construction of an expansion about  
Landau singularities $L(\vec{z})$: $H(\vec{z}) \to 
H(L(\vec{z}))$.
For example, the singular locus $L$ of the Appell function $F_4(z_1,z_2)$ 
is $L = \{ (z_1,z_2) \in \mathbb{C}^2 |z_1 z_2 R(z_1,z_2)  =0 \} \cup L_\infty$
where $R(z_1,z_2) =(1\!-\!z_1\!-\!z_2)^2\!-\!4 z_1z_2$,
and the physically interesting case of an expansion around the singularities 
corresponds to an analytical continuation 
$F_4(z_1,z_2) 
 \to 
 F_4\left(\frac{R(z_1,z_2)}{z_1}, \frac{R(z_1,z_2)}{z_2} \right)$.

A similar problem, the construction of convergent series 
of multiple Mellin-Barnes integrals in different regions of parameters, 
has been analyzed in detail for  
the case of two variables~\cite{tsikh,passare,friot}. 
However, to our knowledge, there are no systematic analyses of the 
relation between these series and the singularities of multiple 
Mellin-Barnes integrals.  

It was understood long ago that there is a one-to-one correspondence 
between the construction of convergent series from Mellin-Barnes 
integrals and the asymptotic expansions; see
Ref.\ \cite{MB:asymptotic} for example. The available software, 
{\em e.g.} Ref.\ \cite{czakon}, allows the construction of
the analytical continuation of a Mellin-Barnes integral in 
the limit when some of the variables $z$ goto to $0$, or $\infty$. 
These are quite useful in the evaluation of Feynman diagrams, but do not 
solve our problem. The current status of the asymptotic expansions is discussed 
in Ref.\ ~\cite{asymptotic1,asymptotic2}. 

Another technique for obtaining a hypergeometric representation 
is the so-called ``Negative Dimensional Integration Method'' (NDIM) 
~\cite{ndim1,ndim2,ndim3,ndim4,ndim5,ndim6}. 
However, it is easy to show~\cite{ndim} that all available results 
follow directly from the Mellin-Barnes integrals~\cite{boos-davydychev}.

For some Feynman diagrams, the hypergeometric representation follows from 
a direct integration of the parametric representation, see 
Ref.\ \cite{somogy,grozin-kotikov,
britto:2015,ablinger:2015,feng1,feng2,yang1,feng3,grozin}.

We also mention that the ``Symmetries of Feynman Integrals''
method~\cite{kol1,kol2,kol3} can also be used to obtain the 
hypergeometric representation for some types of diagrams. 

\subsection{About GKZ and Feynman Diagrams}
\label{GKZ}
There are a number of different though entirely equivalent ways to describe 
hypergeometric functions:
\begin{itemize}
	\item
	as a multiple series;
	\item 
	as a solution of a system of differential equations 
(hypergeometric D-module);
	\item 
	as an integral of the Euler type;
	\item
	as a Mellin-Barnes integral. 
\end{itemize}
In a series of papers, Gel'fand, Graev, Kapranov and Zelevinsky 
~\cite{Gelfand1,Gelfand2,Gelfand3} 
(to mention only a few of their series of papers 
devoted to the systematic development of this approach) 
have developed a uniform  approach to the description of hypergeometric 
functions \footnote{
	The detailed discussion of $A$-functions and their properties 
	is beyond our current consideration.  There are many interesting papers 
	on that subject, including (to mention only a few) 
	Refs.\ \cite{algorithm2,beukers2,review}.}.
The formal solution of the $A$-system is a
so-called multiple $\Gamma$-series having the following form:
$$
\sum_{(l_1, \cdots, l_N) \in \mathbf{L}}
\frac{ z_1^{l_1+\gamma_1} \cdots z_N^{l_N+\gamma_M} }
     {\Gamma(l_1+\gamma_1+1) \cdots \Gamma(l_N+\gamma_N+1) 
}  
\;, 
$$
where $\Gamma$ is the Euler $\Gamma$-function 
and the lattice $ \mathbf{L}$  has rank $d$. 
When this formal series has a non-zero radius of convergence, 
it coincides  (up to a factor) 
with a Horn-type hypergeometric series~\cite{Gelfand3}
(see Section~\ref{Horn-Functions}). 
Any Horn-type hypergeometric function can be written in the form 
of a $\Gamma$-series by applying the reflection formula 
$\Gamma(a+n) = (-1)^n \frac{\Gamma(a) \Gamma(1-a)}{\Gamma(1-a-n)}$. 
Many examples of such a conversion -- 
all Horn-hypergeometric functions of two variables -- 
have been considered in Ref.\ \cite{bod}. 

The Mellin-Barnes representation was beyond Gelfand's consideration. 
It was worked out later by Fritz Beukers~\cite{beukers1};
see also the recent paper~\cite{matsubara}. 
Beukers analyzed the Mellin-Barnes  integral
$$
\int 
\Pi_{i=1}^N 
\Gamma(-\gamma_i - \vec{b}_i \vec{s})
v_i^{\gamma_i+ \vec{b}_i \vec{s}} ds \;, 
$$
and pointed out that,
under the assumption that the Mellin-Barnes integral 
converges absolutely, it satisfies the set
of $A$-hypergeometric equations. 
The domain of convergence 
for the $A$-hypergeometric series and the associated Mellin-Barnes integrals
have been discussed recently in Ref.\ \cite{nilsson}. 

Following Beuker's results, we conclude that any 
Feynman diagram with a generic set of parameters (to guarantee convergence, we 
should treat the powers of propagators as non-integer parameters) 
could be treated as an $A$-function. 
However our analysis has shown that, typically, a real 
Feynman diagram corresponds to an $A$-function with reducible monodromy.  

Let us explain our point of view. 
By studying Feynman diagrams having a one-fold Mellin-Barnes 
representation ~\cite{bkk2009}, we have found that certain Feynman 
diagrams 
($E^q_{1220}, B^2_{1220}, V^q_{1220}, J^q_{1220}$ in the notation of 
Ref.\ \cite{bkk2009}) with powers of propagator equal to one
(the so-called master-integrals) have the following hypergeometric 
structure (we drop the normalization constant for simplicity): 
\begin{eqnarray}
\Phi(n,\vec{1};z)
= 
{}_3F_2(a_1,a_2,a_3;b_1,b_2;z) 
+ 
z^\sigma {}_4F_3(1,c_1,c_2,c_3;p_1,p_2,p_3;z) \;,  
\label{example}
\end{eqnarray}
where the dimension $n$ of space-time \cite{dimreg} is not an integer
and the difference between any two parameters of the hypergeometric 
function also are not integers. 
The holonomic rank of the hypergeometric function 
${}_pF_{p-1}$ is equal to $p$, 
so that the Feynman diagram is a linear combination of two 
series having different holonomic rank.  
What could we say about the holonomic rank of a Feynman diagram $\Phi$? 
To answer that question, let us find the differential equation for 
the Feynman diagram $\Phi(n,\vec{1};z)$ starting from  the
representation Eq.\ (\ref{example}).  This could be done by  the
Holonomic Function Approach~\cite{zeilberger} 
or with the help of a programs
developed by Frederic Chyzak \cite{mgfun}
(it is a MAPLE package)
or by Christoph Koutschan\footnote{
	See Christoph's paper in the present volume.} \cite{HolonomicFunctions} (it is MATHEMATICA package).
We used a private realization of this approach based on ideas from the 
Gr\"obner basis technique. Finally, we obtained the result that  
the Feynman diagram $\Phi$ satisfies the homogeneous differential equation 
of the hypergeometric type of order $4$
with a left-factorizable differential operator of order $1$:  
\begin{equation}
(\theta+A) \left[
(\theta+B_1) (\theta+B_2) (\theta+B_3)
+ z \theta (\theta+C_1) (\theta+C_2) 
\right] \Phi(n,\vec{1};z) = 0 \;, 
\label{MB:F}
\end{equation}
where none of the $B_j$ and $C_a$ are integers
and $\theta = z \frac{d}{dz}$. 

It follows from this differential equation that
the holonomic rank of the Feynman diagram $\Phi$ is equal to $4$, 
and factorization means that the space of solutions
splits into a direct sum of two spaces of dimension 
one and three: $\Phi_{\mbox{dim}}  = 4 = 1 \otimes 3$. 
As it follows from~\cite{beukers3}, the monodromy representation 
of Eq.\ (\ref{MB:F})
is reducible and there is a one-dimensional invariant subspace.
Consequently, 
there are three non-trivial solutions (master-integrals) 
and the one-dimensional invariant subspace corresponds 
to an integral having a Puiseux-type solution 
(expressible in terms of $\Gamma$-functions). 

We pointed out in Ref.\ \cite{bkk2009} that 
a Feynman diagram can be classified by the dimension 
of its irreducible representation.
This can be evaluated by the construction of differential equations 
or by using the dimension of the irreducible representation of the 
hypergeometric functions entering in the r.h.s.\ of 
Eq.\ (\ref{representation}). 
Indeed, in the example considered in Eq.\ (\ref{example}), the 
dimension of the irreducible representation 
${}_4F_3(1,\vec{c};\vec{p};z)$
is equal to $3$ ~\cite{beukers3}, 
so that 
the dimension of the irreducible 
space of the Feynman diagram $\Phi$  is equal to $3$, 
see Eq.(\ref{MB:F}),
and $\Phi$ is expressible via 
the sum of a series (see Eq.~\ref{example}) with an irreducible representation of dimension $3$.

The results of the analysis performed in Ref.\ \cite{bkk2009}, 
are summarized in the following proposition: \\[1em]
{\bf Proposition}:\ {\em A Feynman diagram can be 
treated as a linear combination of Horn-type hypergeometric series  
where each term has equal {\bf irreducible} holonomic rank.}\\[1em]

Examining this new ``quantum number,''
irreducible holonomic rank, 
we discover, and can rigorously prove, 
an extra relation between master-integrals~\cite{KK:extra}.
In many other examples we found a complete agreement between the
results of differential reduction and 
and the results of a reduction based on the IBP 
relations~\cite{IBP1,IBP2}.

The Feynman diagram $J$
considered in Ref.\ \cite{KK:extra}
satisfies the differential equation
\begin{eqnarray}
	&& 
	\left( \theta \!-\! \frac{n}{2} \!+\! I_1 \right)
	\left( \theta \!-\! n \!+\! I_2 \right)
	\left[ 
	\theta 
	\left( \theta \!-\! n \!+\! \frac{1}{2} \!+\! I_3 \right)
	+z
	\left( \theta \!-\! \frac{3n}{2} \!+\! I_4 \right)
	\right] J = 0 \;,
	\label{J} 
\end{eqnarray}
where $I_1,I_2,I_3,I_4$ are integers, $n$ is the dimension of space-time 
and $\theta = z \frac{d}{dz}$.
The dimension of $J$ is $4$ and there are two one-dimensional 
invariant subspaces, corresponding to two first-order differential 
operators:
$
J_{\mbox{dim}}  = 4 = 1 \otimes 1 \otimes 2 \;.
$
Indeed, after integrating twice, we obtained
$$
\left[ 
\theta 
\left( \theta \!-\! n \!+\! \frac{1}{2} \!+\! I_3 \right)
+z
\left( \theta \!-\! \frac{3n}{2} \!+\! I_4 \right)
\right] J = C_1 z^{n/2-I_1} + C_2 z^{n-I_2} \;.
$$
Surprisingly, this simple relation has not been not reproduced 
(as of the end of $2016$)
by any of the powerful programs for the reduction of Feynman diagrams
(see the discussion in Chapter 6 of Ref.\ \cite{KK:sunset}).
	In ~\cite{SS}, it was shown that the extra relation 
	~\cite{KK:extra} could be  deduced from a diagram of more general 
	topology by exploring a new relation
	derived by taking of the derivative with respect to the mass 
	with a subsequent reduction with the help of IBP relations.
	However, it was not shown that the derivative with respect to 
	mass can be deduced from derivatives with respect to momenta, so 
	that the result of Ref.\ \cite{SS} could be considered as an alternative proof 
	that, in the massive case, there may exist an extra relation between 
	diagrams that does not follow from classical IBP relations.

Finally, we have obtained a very simple result~\cite{KK:MB}: 
Eq.\ (\ref{MB:F}) follows directly from the
Mellin-Barnes representation for a Feynman diagram.
(See Section~\ref{Horn-Functions} and Eq.\ (\ref{MB:DE}) for details.)
Based on this observation and on the results of our analysis performed 
in Ref.\ \cite{bkk2009}, and extending the idea of 
the algorithm of Ref.\ \cite{algorithm1}, 
we have constructed a simple and fast algorithm for the algebraic reduction  
of any Feynman diagram having a one-fold Mellin-Barnes integral representation 
to a set of master-integrals without using the IBP relations. 
In particular, our approach and our program cover some types of 
Feynman diagrams with arbitrary powers of propagators considered in 
Refs.\ \cite{mizera1,mizera2,mizera3}.

In a similar manner, one can consider the multiple Mellin-Barnes 
representation of a Feynman diagram~\cite{KK:sunset,hyperdire}. 
In contrast to the one variable case, the factorization of the
partial differential operator is much more complicated.
The dimension of the Pfaff system~\footnote{Rigorously speaking, 
	this system of equations is correct when there is a
	contour  in $\mathbf{C}^n$ that is not changed 
	under translations by an arbitrary unit vector, see 
	Ref.\ \cite{Sadykov}, 
	so that we treat the powers of propagator as parameters.} 
related to the multiple Mellin-Barnes 
integral can be evaluated with the help of a prolongation procedure
(see the discussion in Section~\ref{Horn-Functions}). However, 
in this case, there may exist a Puiseux type solution even for a generic 
set of parameters (see for example Section~\ref{FT:section}).

Exploring the idea~\footnote{The monodromy group is  
	the group of linear transformations of solutions 
	of a system of hypergeometric differential equations under 
	rotations around its singular locus.
	In the case when the monodromy is reducible, there is a 
	finite-dimensional subspace of 
	holomorphic solutions of the hypergeometric system on which 
	the monodromy acts trivially.} 
presented in Ref.\ \cite{beukers3}, one possibility is to 
construct an explicit solution of the invariant subspace 
(see~\cite{KK:sunset} and Section~\ref{FT:section})
and find the dimension of the irreducible representation.  
Our results presented in ~\cite{KK:sunset} were confirmed by another technique 
in Ref.~\cite{bbkp}. 

Let us illustrate the notion of irreducible holonomic rank 
(or an irreducible representation) in an application to Feynman diagrams.
As follows from our analysis of sunset diagrams~\cite{KK:sunset}, 
the dimension of the
irreducible representation of two-loop sunset with three different masses 
is equal to $4$. There is only one hypergeometric function of three 
variables having holonomic rank $4$, the $F_D$ function. 
Then we expect that there is a linear combination of 
four two-loop sunsets and the product of one-loop tadpoles that are
expressible in terms of a linear combination of the $F_D$ functions.  

Another approach to the construction of a GKZ representation 
of Feynman Diagrams
was done recently in the series of papers in Refs.\ \cite{GKZ1,GKZ2,GKZ3}.
Based on the observation made in Ref.\ \cite{Nilsson-Passere}
about the direct relation between $A$-functions and 
Mellin transforms of rational functions, and exploring the 
Lee-Pomeransky representation~\cite{Lee-Pomeransky}, 
the authors studied a different aspect of the GKZ representation 
mainly considering the examples of massless or one-loop diagrams. 
Two non-trivial examples 
have been presented in Ref.~\cite{GKZ2}: the two-loop sunset 
with two different masses and one zero 
mass, which corresponds to a linear combination of two Appell functions 
$F_4$ (see Eq.\ (3.11) in~\cite{JK})~\footnote{It is interesting to note, 
	that on-mass shell $z=1$, this diagram has two Puiseux type solutions
	that do not have analytical continuations.}
and a two-loop propagator with three different masses related 
to the functions $F_C$ of three variables~\cite{berends}. 

A different idea on how to apply the GKZ technique to the analysis of Feynman Diagrams has been
presented in \cite{vanhove} and has received further development in
\cite{klemm1,klemm2}. 

\subsection{One-Loop Feynman Diagrams}
\label{one-loop}
Let us give special attention to one-loop Feynman diagrams. 
In this case, two elegant approaches 
have been developed~\cite{DD,FJT} that allow us to obtain 
compact hypergeometric representations for the master-integrals. 
The authors of the first paper~\cite{DD} explored 
the internal symmetries of the Feynman parametric representation to get a one-fold integral 
representation for one-loop Feynman diagrams (see also ~\cite{bloch-kreimer,n-gon}). 
The second approach~\cite{FJT} 
is based on the solution of difference equations  
with respect to the dimension of space-time~\cite{tarasov:d}
for the one-loop integrals.
In spite of different ideas on the analysis of Feynman diagrams,
both approaches, ~\cite{DD} and ~\cite{FJT},
produce the same results for one-loop propagator and vertex diagrams ~\cite{one-loop:vertex1,one-loop:vertex2,one-loop:vertex3}. 
However, beyond these examples, the situation is less complete: it was 
shown in Ref.\ \cite{FJT} that the off-shell one-loop massive box is expressible in terms of a linear combination of 
$F_S$ Horn-type hypergeometric functions of three variables 
(see also discussions
in Refs.\ \cite{bkm2013,riemann1,riemann2,phan}), 
or in terms of $F_N$ Horn-type hypergeometric functions of three variables ~\cite{davydychev:box} 
(see Section ~\ref{FN:section}).

Recently, it was observed 
~\cite{yangian1,yangian2,yangian3}
that massive conformal Feynman diagrams 
are invariant under a Yangian 
symmetry that allows to get the hypergeometric representation 
for the conformal Feynman diagrams. 
 
\subsection{Construction of $\ep$-expansion}
For physical applications, the construction of the analytical coefficients
of the Laurent expansions of hypergeometric functions around particular values of parameters (integer, half-integer, rational) is necessary. 
Since the analytic continuations of hypergeometric functions 
is still an unsolved problem, the results are written 
in some region of variables in each order of the $\ep$-expansion 
in terms of special functions like classical or multiple polylogarithms, 
~\cite{lewin,harmonic,2dim,BBBL,mpl1,mpl2,mpl3},
and then these functions are analytically continued to another region. For this reason, the analytical properties of special functions 
were
 analyzed in detail~\cite{mpl:properties1,mpl:properties2,mpl:properties3,
	  mpl:properties4,mpl:properties5,panzer2015,PolyLog}.
Also, tools for the numerical evaluation of the corresponding functions are important ingredients 
~\cite{gr1,gr2,vollinga,logsine,maitre1,bonciani2011,maitre2,chaplin,
	 li22,maple,harmonic8,handyG,duhr-tancredi,walden}.  

Each of the hypergeometric function representations  (series, integral, Mellin-Barnes, differential equation) can be used for
the construction of the $\ep$-expansion,
and each of them has some technical advantages or disadvantages
in comparison with the other ones.
The pioneering $\ep$-expansion of the hypergeometric function 
${}_pF_{p-1}$ around $z=\pm 1$ was done by David Broadhurst 
~\cite{david1,david2}. 
The expansion was based on the analysis of multiple series 
and it was interesting from a mathematical point of view~\cite{david3}
as well as for its application to quantum field theory~\cite{david4}.
The integral representation was mainly developed by Andrei Davydychev 
and Bas Tausk~\cite{dt1,dt2}, so that, finally, the all-order 
$\ep$-expansion for the Gauss hypergeometric functions around 
a rational parameter, a case
that covers an important class of diagrams, 
has been constructed~\cite{davydychev} in terms of 
generalized log-sine~\cite{lewin} functions or in term of Nielsen polylogarithms \cite{DK1,DK2}. 

The integral representation was also the starting point for the 
construction of the
$\ep$-expansion of ${}_pF_{p-1}$ hypergeometric functions,~\cite{huber-maitre1,huber-maitre2}, 
and also the $F_1$ \cite{ndim4} and $F_D$ functions 
around integer values of parameters 
~\cite{bogner-brown1,bogner-brown2,panzer2014,bogner-mpl}.

Purely numerical approaches~\cite{NumExp1,pentagon4,NumExp2}
can be applied for arbitrary values of the parameters. 
However, this technique typically does not produce a stable numerical
result in regions around singularities of the hypergeometric functions.

A universal technique which does not 
depend on the order of the differential equation 
is based on the algebra of multiple sums~\cite{nested1,nested2,nested3}.
For the hypergeometric functions\footnote{It was shown  
	in ~\cite{smirnov,tausk} that multiple Mellin-Barnes integrals 
	related to Feynman diagrams could be evaluated analytically/numerically 
	at each order in $\ep$ via multiple sums, without requiring a
	closed expression in terms of Horn-type hypergeometric functions.}
for which the nested-sum algorithms~\cite{nested1} are applicable, 
the results 
of the $\ep$-expansion are automatically obtained in terms of multiple 
polylogarithms.

The nested-sum algorithms~\cite{nested1} have been implemented in a few
packages~\cite{nested1a,nested1b} and allow for the construction of the
$\ep$-expansion of hypergeometric functions ${}_pF_{p-1}$ and 
Appell functions $F_1$ and $F_2$ around integer values of parameters\footnote{
	See also Refs.\ \cite{series1,series2} for an alternative realization.}. 
However, the nested-sum approach fails for 
the $\ep$-expansion of hypergeometric functions 
around rational values of parameters and it is not applicable to 
some specific classes of hypergeometric functions (for example, the $F_4$ 
function, see~\cite{pentagon1}).

In the series of papers~\cite{DK3,kalmykov2004}, the generating function 
technique~\cite{generating1,generating2} has been developed 
for the analytical evaluation of multiple sums. 
Indeed,  the series generated by the $\ep$-expansion of hypergeometric 
functions has the form  
$
\sum_k c(k) z^k\;, 
$
where the coefficients $c(k)$ include only products of the harmonic sums 
$\Pi_{a,b} S_a(k-1) S_b(2k-1)$ and $S_a(k) = \sum_{j=1}^k
\frac{1}{j^a}$. The harmonic sums satisfy the recurrence relations
$$
S_a(k) 
= S_a(k-1) 
+ \frac{1}{k^a}
\quad, \qquad
S_a(2k+1) 
= S_a(2k-1) 
+ \frac{1}{(2k+1)^a}
+ \frac{1}{(2k)^a}\;,
$$ 
so that the coefficients $c(k)$ satisfy 
the first order difference equation\footnote{In general, it could be 
	a more generic recurrence, $\sum_{1=0}^k p_{k+j}(k+j)c(k+j) = r(k)$.}: 
$$
P(k+1) c(k+1) = Q(k) c(k) + R(k) \;, 
$$ where $P$ and $Q$ are polynomial functions 
that can be defined from the original series.
This equation could be converted into 
a first order differential equation for the generating function
$F(z) = \sum_k c(k) z^k$, 
$$
\frac{1}{z} P \left(z \frac{d}{dz} \right) F(z)
- P(1)C(1) z = Q \left(z \frac{d}{dz} \right) F(z)
+ \sum_{k=1} R(k) z^k \;.
$$
One of the 
remarkable properties of this technique, that the non-homogeneous part
of the differential equation, the
function $R(k)$,  has one-unit less depth in contrast to the 
original sums, so that, step-by-step, all sums could be evaluated analytically. 
Based on this technique, all series arising from 
the $\ep$-expansion of hypergeometric functions around 
half-integer values of parameters have been evaluated~\cite{DK3} 
up to weight $4$.
The limits considered were mainly motivated by physical reasons (at $O$(NNLO) only 
functions of weight $4$ are generated) and, in this limit, only one new function~\cite{harmonic} $H_{-1,0,0,1}(z)$ was necessary to introduce.
These results~\cite{DK3} allow us to construct the $\ep$-expansion of 
the hypergeometric functions ${}_pF_{p-1}$
around half-integer values of parameters, see~\cite{JKV2003,MKL:Gauss}.

Other results and theorems relevant for the evaluation of 
Feynman diagrams are related with the appearance of a factor 
$1/\sqrt{3}$ in the $\ep$-expansion of some diagrams~\cite{FK1999} 
expressible in terms of hypergeometric functions  
\footnote{Recent results on the analytical evaluation of inverse binomial 
	sums for particular values of the arguments  
	have been presented in~\cite{binsum1,binsum2,binsum3}.}
were derived in Refs.\ \cite{DK2,KWY2007,KK2010}\footnote{The appearance of 
	$1/\sqrt{3}$ in RG functions in seven loops was quite 
	intriguing~\cite{schnetz1,panzer2015}.}.

Let us consider typical problems arising in this program. We follow   
our analysis presented in Ref.\ \cite{BKK2012}, see also the closely related 
discussion in Ref.\ \cite{abs2018}. 
First, the construction of the difference equation for the coefficient 
functions $c(z)$ is not an easy task~\cite{schneider1,schneider2,schneider3}. 
In the second step, the differential operator(s) coming from the difference 
equation
$P \left(z \frac{d}{dz} \right) -  z  Q \left(z \frac{d}{dz} \right)$
should be factorized   into a product of differential operators of the first 
order, 
$$
P \left(z \frac{d}{dz} \right) -  z  Q \left(z \frac{d}{dz} \right)
= 
\Pi_{k=1} 
\left[ p_{k}(z) \frac{d}{dz}  - q_{k}(z) \right]\;,
$$
where $ p_{k}(z)$ and $q_{k}(z)$ are rational functions.  Unfortunately, the 
factorization of differential operators into irreducible factors is not 
unique~\cite{landau}: 
$$
\left(
\frac{d^2}{dx^2} \!-\! \frac{2}{x} \frac{d}{dx} \!+\! \frac{2}{x^2}
\right)
= 
\left(
\frac{d}{dx} \!-\! \frac{1}{x}
\right)
\left(
\frac{d}{dx} \!-\! \frac{1}{x}
\right)
= 
\left(
\frac{d}{dx} \!-\! \frac{1}{x (1+ax)}
\right)
\left(
\frac{d}{dx} \!-\! \frac{(1+2ax)}{x(1+ax)}
\right) \;, 
$$
where $a$ is a constant. 

However, the 
following theorem is valid (see~\cite{schwarz}):
Any two decompositions of a linear differential operator $L^{(p)}$
into a product (composition) of irreducible linear differential 
operators 
$$
L^{(p)} = 
L_1^{(a_1)}  
L_2^{(a_2)}   
\cdots 
L_m^{(a_m)}  
= 
P_1^{(r_1)}  
P_2^{(r_2)}   
\cdots 
P_k^{(r_k)}
$$
have equal numbers of components $m=k$ 
and the factors $L_j$ and $P_a$ 
have the same order of differential operators:
$L_a=P_j$ (up to commutation). In the application to the $\ep$-expansion 
of hypergeometric functions this problem has been discussed in 
Ref.\ \cite{yost2011}.

After factorization, the iterated integral over 
rational functions (which is not uniquely defined, as seen in 
the previous example) would be generated that in general is not 
expressible in terms of hyperlogarithms. 
Indeed, the solution of the differential equation
$$
	\left[ R_1(z) \frac{d}{dz}  \!+\! Q_1(z) \right] \left[ R_2(z) \frac{d}{dz}  \!+\! Q_2(z) \right] h(z) = F(z) \;.
	\label{de}
$$
has the form
$$
	h(z) = 
	\int^{z} \frac{dt_3}{R_2(t_3)} \left[ \exp^{-\int_0^{t_3}\frac{Q_2(t_4)}{R_2(t_4)} dt_4} \right] 
	\int^{t_3} \frac{dt_1}{R_1(t_1)} \left[ \exp^{-\int_0^{t_1}\frac{Q_1(t_2)}{R_1(t_2)} dt_2} \right] F(t_1) \;. 
$$
    From this solution it follows~\cite{KK2010B}
	that the following conditions are enough to 
	convert the iterated integral into hyperlogarithms:
	there are new variables $\xi$ and $x$ so that  
	\begin{eqnarray}
		&& 
		\int^z \frac{Q_i(t)}{R_i(t)} dt  =  \ln \frac{M_i(\xi)}{N_i(\xi)} 
		\Rightarrow
		\left. \frac{dt}{R_i(t)} \right|_{t = t(\xi)} 
		\frac{N_i(\xi)}{M_i(\xi)}  =  dx \frac{K_i(x)}{L_i(x)} \;, 
		\nonumber 
	\end{eqnarray}
	where $M_i,N_i,K_i,L_i$ are polynomial functions.

The last problem is related to the Abel-Ruffini theorem:  
the polynomial is factorizable into a product of its primitive roots, 
but there are not solutions in radicals for 
polynomial equations of degree five or more. 
The last problem got a very elegant solution by the introduction 
of cyclotomic polylogarithms~\cite{nested3}, with the integration over 
irreducible cyclotomic polynomials $\Phi_n(x)$.
The first two irreducible polynomials (see Eqs.\ $(3.3)-(3.14)$ in~\cite{nested3}) 
are  $\Phi_7$ and $\Phi_9$
(the polynomial of order $6$). 
Two other polynomials of order $4$, 
$\Phi_5$ and $\Phi_{10}$: 
$(x^4 \pm x^3 + x^2 \pm x +1)$, have 
non-trivial primitive roots. But up to now, all these polynomials were not 
generated by Feynman diagrams. Surprisingly, 
by increasing the number of loops or number of scales, 
other mathematical structures are generated.~\cite{bs1,bs2}.
Detailed analyses of properties of the new functions have been presented in 
Refs.\ \cite{abs2013,abs2014} and automated by Jakob 
Ablinger~\cite{ablinger1,ablinger2,ablinger3}. 
The problem of integration over algebraic functions 
(typically square roots of polynomials) 
was solved by the introduction of a new type of functions,
~\cite{Bonciani}, intermediate between multiple and elliptic polylogarithms.

The series expansion is not very efficient for the construction 
of the $\ep$-expansion, since the number of series increases with the 
order of the $\ep$-expansion and increases the complexity of the individual 
sums.
Let us recall that the Laurent expansion of a hypergeometric function 
contains a linear combination of multiple sums. 
From this point of view, the construction of the analytical coefficients of
the $\varepsilon$-expansion of a hypergeometric function
can be carried out independently
of existing analytical results for each individual multiple sum. 
The ``internal'' symmetry of a Horn-type hypergeometric function is 
uniquely defined by the corresponding system of differential equations. 
While exploring this idea, a new algorithm was presented in 
Refs.\ \cite{kwy2006,kwy2007}, based on factorization, looking for a 
linear parametrization and direct iterative solution 
of the differential equation for a hypergeometric function.
This approach allows the construction of the analytical coefficients of the 
$\varepsilon$-expansion of a hypergeometric 
function, as well as obtaining analytical expressions 
for a large class of multiple series 
without referring to the algebra of nested sums.

Based on this approach, the all-order $\varepsilon$-expansion of the Gauss 
hypergeometric function around half-integer and rational values of parameters
has been constructed~\cite{kwy2006,kk2008}, so that the first $20$ coefficients 
around half-integer values of parameters, the $12$ coefficients 
for $q=4$ and $10$ coefficients for $q=6$
have been generated already in 2012~\footnote{The results 
	have been written in 
	terms of hyperlogarithms of primitive $q$-roots of unity.}. 
Another record is the generation of $24$ coefficients for the
Clausen hypergeometric function ${}_3F_2$ around integer values of 
parameters, relevant for the analysis~\footnote{The results of 
	~\cite{david5} were relevant for the reduction of multiple zeta values
	to the minimal basis.} 
performed in ~\cite{boels}.
To our knowledge, at the present moment, this remains the fastest and most 
universal algorithm.

Moreover, it was shown in Refs.\ \cite{kwy2006,kwy2007}, 
that when the coefficients of the $\varepsilon$-expansion of a hypergeometric 
function are expressible in terms of multiple polylogarithms, there is a set 
of parameters (not uniquely defined) such that, 
at each order of $\varepsilon$, the coefficients of the $\varepsilon$-expansion 
include multiple polylogarithms of a single uniform weight. 
A few years later, this property was established 
not only for hypergeometric functions, but for Feynman Diagrams
~\cite{Henn:1}.

A multivariable generalization~\cite{BKK2012} 
of the algorithm of Refs.\ \cite{kwy2006,kwy2007} has been described.
The main difference with respect to the case of one variable is the 
construction of a system of differential equations of triangular form
 to avoid the appearance of elliptic functions.
As a demonstration of the validity of the algorithm, the first few coefficients 
of the $\varepsilon$-expansion of the Appell hypergeometric functions 
$F_1,F_2,F_3$ and $F_D$ around integer values 
of parameters have been evaluated analytically~\cite{bkm2013}. 

The $\varepsilon$-expansion of the hypergeometric 
functions $F_3$ and $F_D$ 
are not covered by the nested sums technique or its generalization. 
The  differential equation
technique can be applied to the construction of analytic coefficients of 
the $\varepsilon$-expansions of hypergeometric functions of several variables 
(which is equivalent to the multiple series of several variables)
around any rational values of parameters via direct solution of the 
linear systems of differential equations. 

The differential equation approach~\cite{kwy2006,kwy2007}
allows us to analyze arbitrary sets of parameters simultaneously
and to construct the solution in terms of iterated integrals, 
but for any hypergeometric function
the Pfaff system of differential equations should be constructed.
That was the motivation for creation of the package(s) (the HYPERDIRE 
project) ~\cite{hyperdire} for the manipulation of the parameters of 
Horn-type hypergeometric functions of several variables. 
For illustration, we describe in detail how it works in the application 
to the $F_3$ hypergeometric function in Section~\ref{F3:general}.

Recently, a new technique~\cite{abreu1}
for the construction of the $\ep$-expansion 
of Feynman diagrams~\cite{abreu2}
as well as for hypergeometric functions has been presented~\cite{abreu3}.
It is based on the construction of a coaction~\footnote{An interesting 
	construction of the coaction for the Feynman graph has been presented 
	recently in Ref.\ \cite{Kreimer2020}.} 
of certain hypergeometric functions. The structures of the $\ep$-expansion of 
the Appell hypergeometric functions $F_1,F_2,F_3$ and $F_4$ as well as $F_D$
(for the last function $F_D$ see also the discussion in Ref.\ \cite{brown-dupont}) 
around integer values of parameters 
are in agreement with our analysis and partial results presented in
Refs.\ \cite{BKK2012} and ~\cite{bkm2013}. However, the 
structure of the $\ep$-expansion around rational values of parameters 
has not been discussed in~\cite{abreu3}, nor in ~\cite{brown-dupont}.

\section{Horn-type hypergeometric functions}
\label{Horn-Functions}
\subsection{Definition and system of differential equations}
\label{Horn:defintion}
The study of solutions of linear partial 
differential equations (PDEs) of several variables in terms of multiple series,
{\em i.e.} a multi-variable generalization of the Gauss hypergeometric 
function~\cite{Gauss}, began long ago~\cite{Lauricella}.

Following the Horn definition~\cite{horn}, a multiple series is called a 
``Horn-type hypergeometric function,'' if, about the point $\vec{z}=\vec{0}$, 
there is a series representation
\begin{equation}
H(\vec{z}) = \sum_{\vec{m}} C(\vec{m}) \vec{z}^{\vec{m}},
\label{def}
\end{equation}
where 
$
\vec{z}^{\vec{m}} = z_1^{m_1} \cdots z_r^{m_r}
$
for any integer multi-index
$\vec{m} = (m_1, \cdots, m_r)$, 
and the ratio of two coefficients can be represented as a ratio of
two polynomials:
\begin{eqnarray}
\frac{C(\vec{m}+\vec{e}_j)}{C(\vec{m})}  
=  \frac{P_j(\vec{m})}{Q_j(\vec{m})} 
\;,
\label{horn}
\end{eqnarray}
where
$
\vec{e}_j
$
denotes the unit vector with unity in its $j^{\rm th}$ entry, 
$                   
\vec{e}_j = (0,\cdots,0,1,0,\cdots,0).  
$
The coefficients $C(\vec{m})$ of such a series can be expressed as 
products or ratios of Gamma-functions (up to some factors 
irrelevant for our consideration)~\cite{Ore:Sato:1,Ore:Sato:2}:
\begin{eqnarray}                                              
	C(\vec{m})=
\frac{
	\prod\limits_{j=1}^p
	\Gamma\left(\sum_{a=1}^r \mu_{ja}m_a+\gamma_j \right)}
{
	\prod\limits_{k=1}^q
	\Gamma\left( \sum_{b=1}^r \nu_{kb}m_b+\sigma_k \right)
} \;,
\label{ore}
\end{eqnarray}
where
$                                                             
\mu_{ja}, \nu_{kb}, \sigma_j,\gamma_j \in \mathbb{Z}
$
and $m_a$ are elements of ${\vec{m}}$.

The Horn-type hypergeometric function, Eq.~(\ref{horn}),  
satisfies the following system of differential equations:
\begin{equation}
0 =
L_j (\vec{z})
H(\vec{z})
=
\left[
Q_j\left(
\sum_{k=1}^r z_k\frac{\partial}{\partial z_k}
\right)
\frac{1}{z_j}
-
P_j\left(
\sum_{k=1}^r z_k\frac{\partial}{\partial z_k}
\right)
\right]
H(\vec{z}) \;,
\label{diff}
\end{equation}
where $j=1, \ldots, r$.
Indeed, 
\begin{eqnarray}
&& 
Q_j\left(
\sum_{k=1}^r z_k\frac{\partial}{\partial z_k}
\right)
\frac{1}{z_j} \sum_{\vec{m}} C(\vec{m}) \vec{z}^{~\vec{m}}
=  \sum_{\vec{m}} Q_j(\vec{m}) C(\vec{m}\!+\!\vec{e}_j) \vec{z}^{~\vec{m}}
\nonumber \\ && 
= 
\sum_{\vec{m}} P_j(\vec{m}) C(\vec{m}) \vec{z}^{~\vec{m}}
=
P_j\left(
\sum_{k=1}^r z_k\frac{\partial}{\partial z_k}
\right)
\sum_{\vec{m}}  C(\vec{m}) \vec{z}^{~\vec{m}} \;.
\nonumber 
\end{eqnarray}
The degrees of the polynomials $P_i$ and $Q_i$ 
are $p_i$ and $q_i$, respectively.
The largest of these, $r=\max \{p_i,q_j \}$, is called the order of 
the hypergeometric series. 
To close the system of differential equations, the 
{\it prolongation procedure} 
should be applied: by
applying the differential operator $\partial_i$ to $L_j$
we can convert the system of linear PDEs with polynomial 
coefficients into Pfaff form (for simplicity, we assume that system is closed): 

\begin{eqnarray}
L_j H(\vec{z}) = 0 
& \Rightarrow & 
\Biggl\{
d \omega_i(\vec{z}) = \Omega_{ij}^k(\vec{z}) \omega_j(\vec{z}) dz_k  \;,
\quad 
d \left[ d \omega_i(\vec{z}) \right] = 0 
\Biggr\} \;.
\label{pfaff}
\end{eqnarray}

Instead of a series representation, one can 
use a Mellin-Barnes integral representation 
(see the  discussion in ~\cite{KK:MB}). Indeed,  
the multiple Mellin-Barnes representation for a Feynman diagram 
could be written in the form in Eq.\ (\ref{MB}).
Let us define the polynomials $P_i$ and $Q_i$ as 
\begin{equation}
\frac{P_i(\vec{t})}{Q_i(\vec{t})} = \frac{\phi(\vec{t}+e_i)}{\phi(\vec{t})} \;.
\end{equation}
The integral (\ref{MB}) then satisfies the system of linear differential 
equations (\ref{diff})
\begin{eqnarray}
\left. Q_i(\vec{t}) \right|_{t_j \to \theta_j }\frac{1}{z_i} \Phi({\bf A},
	\vec{B};{\bf C}, \vec{D};\vec{z})  
	= \left. P_i(\vec{t}) \right|_{t_j \to \theta_j } \Phi({\bf A},
	\vec{B};{\bf C}, \vec{D};\vec{z})  \;,
\label{MB:DE}
\end{eqnarray}
where $\theta_i =  z_i\frac{d}{dz_i}$.
Systems of equations such as Eq. (\ref{MB:DE}) are left ideals in 
the Weyl algebra of linear differential operators with polynomial coefficients.

\subsection{Contiguous relations}
\label{Horn:contiguous}
Any Horn-type hypergeometric function is a function of two types of variables, 
{\it continuous} variables, $z_1,z_2, \cdots, z_r$ and 
{\it discrete} variables:
$\{ J_a \}:= \{\gamma_k,\sigma_r \}$, 
where the latter can change by integer numbers 
and are often referred to as the {\it parameters} of the hypergeometric 
function. 

For any Horn-hypergeometric function, 
there are linear differential operators 
changing the value of the discrete variables by one unit. 
Indeed, let us consider a multiple series defined by 
Eq.~(\ref{def}).

Two hypergeometric functions $H$
with sets of parameters shifted by unity,
$H(\vec{\gamma}+\vec{e_c};\vec{\sigma};\vec{z})$ and
$H(\vec{\gamma};\vec{\sigma};\vec{z})$,
are related by a linear differential operator:
\begin{eqnarray}
H(\vec{\gamma}+\vec{e_c};\vec{\sigma};\vec{z})
& = &
\left ( \sum_{a=1}^r \mu_{ca} 
z_a \frac{\partial}{\partial z_a}+\gamma_c \right)
H(\vec{\gamma};\vec{\sigma};\vec{z})
\;.
\label{do1}
\end{eqnarray}
Similar relations also exist for the lower parameters:
\begin{eqnarray}
H(\vec{\gamma};\vec{\sigma}-\vec{e}_c;\vec{z})
& = &
\left(
\sum_{b=1}^r
\nu_{cb} z_b \frac{\partial}{\partial z_b} \!+\! \sigma_c \!-\! 1
\right)
H(\vec{\gamma};\vec{\sigma};\vec{z})
\;.
\label{do2}
\end{eqnarray}
Let us rewrite these relations in a symbolic form: 
\begin{eqnarray}
R_{K}(\vec{z})\frac{\partial }{\partial \vec{z}_K} H(\vec{J};\vec{z}) 
= H(\vec{J} \pm e_K; \vec{z}) \;,
\label{direct}
\end{eqnarray}
where $R_{K}(\vec{z})$ are polynomial (rational) functions.

In Refs.\ \cite{algorithm1,algorithm2} it was shown that there is an 
algorithmic construction of inverse linear differential operators: 
\begin{eqnarray}
B_{L,N}(\vec{z})\frac{\partial^L }{\partial \vec{z}_N} 
\left( R_{K}(\vec{z})\frac{\partial }{\partial \vec{z}_K} \right) 
H(\vec{J};\vec{z}) 
\equiv 
B_{L,N}(\vec{z})\frac{\partial^L }{\partial \vec{z}_N} H(\vec{J} \pm e_K; \vec{z}) 
= 
H(\vec{J};\vec{z}) 
\;.
\label{inverse}
\end{eqnarray}
Applying the direct or inverse differential operators
to the hypergeometric function
the values of the parameters can be changed by an arbitrary integer: 
\begin{equation}
S(\vec{z}) H(\vec{J}+\vec{m}; \vec{z})
= 
\sum_{j=0}^r S_j(\vec{z}) 
\frac{\partial^j}{\partial \vec{z}} H(\vec{J}; \vec{z}) \;,
\label{reduction}
\end{equation}
where $\vec{m}$ is a set of integers, $S$ and $S_j$ are polynomials 
and $r$ is the holonomic rank (the number of linearly independent solutions) 
of the system of differential equations Eq.~(\ref{diff}).
At the end of the reduction, the differential operators acting on the 
function $H$ can be  replaced by a linear combination of the function 
evaluated with shifted parameters. 

We note that special considerations are necessary when the system of 
differential operators,  Eq.~(\ref{diff}), has a
Puiseux-type solution (see Section \ref{FT:section}). 
In this case, the prolongation procedure gives 
rise to the Pfaffian form, but
this set of differential equations is not enough to construct 
the inverse operators~\cite{BK}, so that new 
differential equations should be introduced.
In the application to the Feynman diagrams, this problem 
is closely related with obtaining new relations between 
master integrals, see Ref.~\cite{KK:extra} for details.
In the Section ~\ref{FT:section} we present an example of the
Horn-type hypergeometric equation of second order 
of three variables having a Puiseux-type  solution. 

Another approach to the reduction of hypergeometric functions 
is based on the explicit algebraic solution of the contiguous relations, 
see the discussion in Ref.\ \cite{Schlosser}. 
This technique is applicable in many particular cases, including  
${}_2F_1, {}_3F_2$, and the Appell functions $F_1,~F_2,~F_3,~F_4$ 
(see the references in Ref.\ \cite{acat}), and there is 
a general expectation that it could be solved for any Horn-type hypergeometric function. However, to our knowledge, 
nobody has analyzed the algebraic reduction in the application to  
general hypergeometric functions having a Puiseux-type solution.  

The multiple Mellin-Barnes integral
$\Phi$ defined by Eq.~(\ref{MB}) satisfies similar 
differential contiguous relations: 
\begin{eqnarray}
\Phi({\bf A}, \vec{B} \!+\! e_a; {\bf C}, \vec{D};\vec{z}) 
& = & 
\left(\sum_{i=1}^m A_{ai} \theta_i \!+\! B_a \right) 
\Phi({\bf A},\vec{B}; {\bf C}, \vec{D};\vec{z}) \;,
\nonumber \\  
\quad 
\Phi({\bf A}, \vec{B}; {\bf C}, \vec{D} \!-\! e_b;\vec{z}) 
& = &  
\left(\sum_{j=1}^r C_{bj} \theta_j \!+\! D_{b} \right)
\Phi({\bf A},\vec{B}; {\bf C}, \vec{D};\vec{z}) 
\;,
\label{step-up-down}
\end{eqnarray}
so that  
the original diagram may be explicitly reduced to a set of basis functions
without examining the IBP relations~\cite{IBP1,IBP2}.
A non-trivial example of this type of reduction beyond IBP relations 
has been presented in Ref.\ \cite{KK:extra} 
(see also the discusion in Chapter 6 of Ref.\ \cite{KK:sunset}).

\section{Examples}
\label{Example}
\subsection{Holonomic rank \& Puiseux-type solution} 
In addition to the examples presented previously in our series of 
publications, we present here a few new examples.

\subsubsection{Evaluation of holonomic rank: the hypergeometric 
	function $F_N$}
\label{FN:section}
The Lauricella-Saran hypergeometric function of three variables $F_N$ 
 is defined about the point $z_1=z_2=z_3=0$ by
\begin{eqnarray}
&& 
F_N(a_1,a_2,a_3;b_1,b_2; c_1,c_2; z_1,z_2,z_3)
\nonumber \\ && 
= \sum_{m_1,m_2,m_3=0}^\infty
\left[ 
\Pi_{j=1}^3 (a_j)_{m_j} \frac{z_j^{m_j}}{m_j!}
\right] 
\frac{(b_1)_{m_1+m_3} (b_2)_{m_2}}      
{(c_1)_{m_1} (c_2)_{m_2+m_3}} 
\;. 
\label{FN:series}
\end{eqnarray}
This function is related to one-loop box diagrams in an 
arbitrary dimension considered by Andrei Davydychev~\cite{davydychev:box}. 

Following the general algorithm~\cite{bkm2013}, 
the following result is easily derived: 
\begin{theorem}
	For generic values of the parameters, 
	the holonomic rank of the function $F_N$ 
	is equal $8$. 
\end{theorem}
In this way, for generic values of parameters, 
the result of differential reduction, Eq.\ (\ref{reduction}), 
have the following form: 
$$
S(\vec{z}) F_N(\vec{J}+\vec{m}; \vec{z})
= 
\left[ S_0
+ S_i \sum_{j=1}^3 \theta_j 
+ \sum_{\substack{i,j=1 \\ i<j}}^3  S_{i,j} \theta_i \theta_j 
+ S_{123} ~\theta_1 \theta_2 \theta_3 
\right] F_N(\vec{J}; \vec{z}) \;,
$$
where $\theta_i =  z_i\frac{d}{dz_i}$
and inverse differential operators can be easily constructed
(Note that  ``easy'' does not mean that these operators have a simple form, 
see Ref.\ \cite{BK:FC}). 

\begin{theorem}
The system of differential equations defined by the series 
(\ref{FN:series})
is reducible when the one of the following combinations of parameters 
is an integer: 
	\begin{eqnarray}
	&& 
	\left\{ 
	a_1, a_2, a_3, b_1, b_2,
	\right\}
	\in \mathbb{Z}  \;, 
	\nonumber \\ &&  
	\left\{
	a_2\!-\!c_1\!-\!c_2\!+\!b_1, \quad 
	a_2\!-\!c_2\!+\!b_1, \quad  
	b_2\!-\!c_1\!-\!c_2\!+\!b_1, \quad  
	b_2\!-\!c_2\!+\!b_1 \right\}
	\in \mathbb{Z}  
	\end{eqnarray}
	\label{FN:monodromy}
\end{theorem}
When one or more conditions of the Theorem \ref{FN:monodromy}
are valid, the number of independent differential equations 
describing the function $F_N$ reduces 
and some additional analysis is necessary (see for example 
Ref. ~\cite{KK:sunset}) to evaluate  
the value of irreducible holonomic rank.

\subsubsection{Puiseux-type solution:  
hypergeometric function $F_T$}
\label{FT:section}
The Lauricella-Saran hypergeometric function of three variables $F_T$  
is defined about the point $z_1=z_2=z_3=0$ by
\begin{eqnarray}
&& 
F_T(a_1;a_2;b_1,b_2; c; z_1,z_2,z_3)
\nonumber \\ && 
= 
\sum_{m_1,m_2,m_3=0}^\infty
\frac{(a_1)_{m_1} (a_2)_{m_2+m_3} (b_1)_{m_1+m_3} (b_2)_{m_2}}      
{(c)_{m_1+m_2+m_3}} 
\frac{z_1^{m_1}  z_2^{m_2}  z_3^{m_3}}{m_1! m_2! m_3!} 
\;.
\nonumber \\ && 
\label{FT:series}
\end{eqnarray}
In this case, the differential operators, Eq.~(\ref{diff}), are the following: 
\begin{subequations}
	\begin{eqnarray}
	L_1 F_T: && 
	\quad 
	\theta_1 \left(c \!-\! 1 \!+\! \sum_{j=1}^3 \theta_j \right) F_T =  
	z_1 \left(a_1 \!+\! \theta_1 \right) \left(b_1 \!+\! \theta_1 \!+\! \theta_3 \right) F_T \;,
	\\  
	L_2 F_T: && 
	\quad 
	\theta_2 \left(c-1 \!+\! \sum_{j=1}^3 \theta_j \right) F_T =  
	z_2 \left(a_2 \!+\! \theta_2 \!+\! \theta_3 \right) \left(b_2 \!+\! \theta_2 \right) F_T
	\;, 
	\\
	L_3 F_T: && 
	\quad 
	\theta_3 \left(c-1 \!+\! \sum_{j=1}^3 \theta_j \right) F_T =  
	z_3 \left(a_2 \!+\! \theta_2 \!+\! \theta_3 \right) \left(b_1 \!+\! \theta_1 \!+\! \theta_3 \right) F_T
	\;, 
	\end{eqnarray}
	\label{FT:diff}
\end{subequations}
where 
$
F_T = F_T(a_1;a_2;b_1,b_2; c; z_1,z_2,z_3) \;.
$

Let us introduce the function 
\begin{equation}
\Phi_T = \frac{z_1^{1-c+a_2} z_2^{1-c+b_1}  }
{ z_3^{1-c+b_1+a_2} } \;.
\label{PhiT}
\end{equation}
It is easy to check that 
$$
L_1 \Phi_T = L_2 \Phi_T = L_3 \Phi_T = 0 \;.
$$

\begin{theorem}
	The system of differential equations defined by Eq.~(\ref{FT:diff}) 
has a Puiseux-type  solution: 
	\begin{equation}
	\Phi_T = \frac{z_1^{1-c+a_2} z_2^{1-c+b_1}  }
	{ z_3^{1-c+b_1+a_2} } \;.
	\end{equation}
	\end{theorem}
In particular, to construct the inverse contiguous relations 
for the function $F_T$, one extra differential equation should be added. 

For completeness, we also note the following result:
\begin{theorem}
	The monodromy group of the system of differential equations 
defined by Eq.~(\ref{FT:diff}) 
is reducible when the one of the following combinations of parameters is 
an integer: 
$$
		\left\{ 
		a_1, a_2, b_1, b_2, \quad 
		c\!-\!b_1\!-\!b_2, \quad 
		c\!-\!a_1\!-\!a_2
		\right\} 
		\in \mathbb{Z}   \;.
$$
\end{theorem}

A Puiseux-type solution for the hypergeometric differential equation
of two variables was established by Erdelyi~\cite{erdelyi:1950} still in the 50's and has been 
analyzed in detail in ~\cite{sadykov} in the framework of the GKZ approach.

\subsection{
Construction of the $\ep$-expansion via differential equations: the Appell Function $F_3$:}
\label{F3:general}
To explain the technical details of our algorithm, 
let us analyze and construct 
the $\ep$-expansion 
for the Appell hypergeometric function $F_3$. The preliminary results 
have been presented in ~\cite{yost2011} and ~\cite{BKK2012}. 

\subsubsection{Notations}
Let us consider the Appell hypergeometric function $F_3$ defined about
 $x=y=0$ as

\begin{eqnarray}
\omega_0 \equiv F_3(a_1, a_2, b_1, b_2, c; x,y)
=
\sum_{m=0}^\infty
\sum_{n=0}^\infty
\frac{(a_1)_{m} (a_2)_{n} (b_1)_m (b_2)_n}{(c)_{m+n} }
\frac{x^m}{m!}
\frac{y^n}{n!} \;,
\label{definition:f3}
\end{eqnarray}
It is symmetric with respect to simultaneous exchange
\begin{eqnarray}
a_1 \Leftrightarrow a_2 \;, \quad
b_1 \Leftrightarrow b_2 \;, \quad
x  \Leftrightarrow y  \;.
\label{F3:symmetry}
\end{eqnarray}
so that
$
F_3(a_1,a_2,b_1,b_2,c;x,y)
=
F_3(a_2,a_1,b_2,b_1,c;y,x) \;.
$

This function satisfies four differential equations
(see Section 3.4 in ~\cite{hyperdire}): 
\begin{subequations}
	\begin{eqnarray}
	(1\!-\!x) \theta_{xx}  \omega_0
	& = &  - \theta_{xy} \omega_0
	+
	\left[
	(a_1 \!+\! b_1) x \!-\! (c \!-\! 1)
	\right] \theta_x \omega_0
	\!+\! x a_1 b_1 \omega_0 \;,
	\\
	(xy\!-\!x\!-\!y) \theta_{xxy} \omega_0 & = &
	\left[
	(1\!-\!y)(a_1 \!+\! b_1)x \!-\! y (a_2 \!+\! b_2 \!+\! 1 \!-\! c)
	\right] \theta_{xy} \omega_0
	\nonumber \\ &&
	+ (1\!-\!y) x a_1 b_1 \theta_y \omega_0
	\!-\! y a_2 b_2 \theta_x \omega_0 \;,
	\end{eqnarray}
	\label{F3:system}
\end{subequations}
and two other equations follow from  Eqs.~(\ref{F3:system}) and the symmetry 
relation (\ref{F3:symmetry}). 
Eqs.\ (\ref{F3:system}) can be written in projective space with homogeneous coordinates 
on ${\bf P}_2(C)$ with $x = X/Z$ and $y = Y/Z$.  
We want also to remark that the system of differential equations
for the Appell hypergeometric function 
$F_2(u,v)$ coincides with the system Eq.\ (\ref{F3:system})
by a redefinition $u=1/x,v=1/y$. This result follows also from 
analytic continuation of the Mellin-Barnes representation for 
Appell's functions $F_2$ and $F_3$.

\subsubsection{One-fold iterated solution}
Assuming the most general form of the parameters
$$
a_j =  \frac{p_j}{q} \!+\! a_1 \varepsilon \;, \quad 
b_j =  \frac{r_j}{q} \!+\! b_j \varepsilon \;, \quad  
c =  1 - \frac{p}{q} + c \varepsilon \; . 
$$	
where $p_i,r_i,p,q$ are integers, 
and writing the $\varepsilon$-expansion for 
the functions as 
\begin{eqnarray}
\omega_j = \sum_{k=0}^\infty \omega_j^{(k)} \varepsilon^k \;, \quad j = 0,1,2,3,
\end{eqnarray}
we  obtain the full system of differential equations
(under the conditions $p_j r_j = 0 \;, j=1,2$):
\begin{subequations}
	\begin{eqnarray}
	&& 
	\left[ (1\!-\!x) \frac{d}{dx} 
	\!-\! \frac{s_1}{q} 
	\!-\! \frac{1}{x} \frac{p}{q} 
	\right]
	\omega_1^{(r)}
		= - \frac{1}{x} \omega_3^{(r)}
	\!+\!
	\left[
	(a_1 \!+\! b_1) \!-\! \frac{c}{x}
	\right] \omega_1^{(r-1)}
\nonumber \\ && \hspace{5mm}
	+
	\left[ 
	\frac{a_1 r_1\!+\!b_1 p_1}{q} 
	\right]  \omega_0^{(r-1)}
	\!+\! a_1 b_1 \omega_0^{(r-2)} 
	\;,
	\label{f3:1}
	\\
	&& 
	\left[ (1\!-\!y) \frac{d}{dy} - \frac{s_2}{q} - \frac{1}{y} \frac{p}{q} \right] \omega_2^{(r)}
	=  
	- \frac{1}{y} \omega_3^{(r)}
	+
	\left[
	(a_2 \!+\! b_2) \!-\! \frac{c}{y}
	\right] \omega_2^{(r-1)}
	\nonumber \\ && \hspace{5mm}
	+ 
	\left[ 
	\frac{a_2 r_2+b_2 p_2}{q}  
	\right] \omega_0^{(r-1)}
	\!+\! a_2 b_2s \omega_0^{(r-2)}
	\;,
	\\ [1cm]
	&& 
	\left[
	(xy\!-\!x\!-\!y) \frac{d}{dx} 
	- (1-y) \frac{s_1}{q}
	+ \frac{y}{x} \frac{(s_2 + p)}{q}
	\right]
	\omega_3^{(r)}  
	\nonumber \\ && \hspace{5mm}
	=
	\left[
	(1\!-\!y)(a_1 \!+\! b_1) \!-\! \frac{y}{x} (a_2 \!+\! b_2  \!-\! c)
	\right] \omega_3^{(r-1)}
	\nonumber \\ && \hspace{5mm}
	+ 
	(1\!-\!y)
	\left[ 
	\frac{a_1 r_1 \!+\! b_1 p_1}{q}  \omega_2^{(r-1)} 
	+ a_1 b_1  \omega_2^{(r-2)}
	\right]
	\nonumber \\ && \hspace{5mm}
	\!-\! \frac{y}{x}  
	\left[
	\frac{a_2 r_2 \!+\! b_2 p_2}{q}  \omega_1^{(r-1)} 
	\!+\! a_2 b_2  \omega_1^{(r-2)} 
	\right]
	\;,
	\nonumber \\
	\label{f3:2}
	\\ [1cm]
	&& 
	\left[
	(xy\!-\!x\!-\!y) \frac{d}{dy} 
	- (1-x) \frac{s_2}{q}
	+ \frac{x}{y} \frac{(s_1 + p)}{q}
	\right]
	\omega_3^{(r)} 
	\nonumber \\ &&
	= 
	\left[
	(1\!-\!x)(a_2 \!+\! b_2) 
	\!-\! 
	\frac{x}{y} (a_1 \!+\! b_1  \!-\! c)
	\right] \omega_3^{(r-1)} 
	\nonumber \\ &&
	+ (1\!-\!x) 
	\left[ 
	\frac{a_2 r_2 + b_2 p_2}{q} \omega_1^{(r-1)} 
	+ a_2 b_2 \omega_1^{(r-2)}  
	\right]
	\nonumber \\ && \hspace{5mm}
	\!-\! 
	\frac{x}{y}
	\left[
	\frac{a_1 r_1 + b_1 p_1}{q} \omega_2^{(r-1)} 
	+ a_1 b_1 \omega_2^{(r-2)} 
	\right]
	\;,
	\nonumber \\
	\end{eqnarray}
	\label{F3:system:expanded}
\end{subequations}
where we have introduced new notations: 
\begin{equation}
s_1  = p_1 + r_1 \;, \quad 
s_2 = p_2 + r_2  \;.
\label{s1s2}
\end{equation}
We want to mention that this system of differential equations does not have 
the $\ep$-form {\em a-la} Henn's form in ~\cite{Henn:1,Henn:2}. 
However, 
the system of equations Eq.~(\ref{F3:system:expanded})
can be straightforwardly solved iteratively. 
Let us redefine functions
$
\omega_1, \omega_2, \omega_3, 
$
as follows:
\begin{equation}
\omega_{1}^{(r)} = \theta_{x} \omega_0 =  h_1(x) \phi_{1}^{(r)} \;, 
\quad 
\omega_2^{(r)}  = \theta_y \omega_0 =  h_2(y) \phi_2^{(r)} \;, 
\end{equation}
where $h_{1,2}(x)$ are new functions, defined as
\begin{equation}
h_1(x) =  
\sigma_1
\left[ 
\frac{x^p}{(x-1)^{s_1+p}}
\right]^\frac{1}{q} 
\;, 
\quad 
h_2(y)  =   
\sigma_2
\left[ 
\frac{y^p}{(y-1)^{s_2+p}}
\right]^\frac{1}{q} 
\;, 
\label{h}
\end{equation}
and 
\begin{eqnarray}
\omega_{3}^{(r)} & = & \theta_{x} \theta_y \omega_0^{(r)} = H(x,y) \phi_{3}^{(r)} \;, 
\label{H}
\end{eqnarray}
with
\begin{eqnarray}
H(x,y) & = &  
\sigma_3
\left[ 
\frac{x^{s_2+p}y^{s_1+p}}
{(xy-x-y)^{s_1+s_2+p}}
\right]^\frac{1}{q} 
\;, 
\end{eqnarray}
where $\sigma_j, j=1,2,3$ are some normalization constants.

Substituting it to the original system, Eq.~(\ref{F3:system:expanded}), we have 
(for completeness we show all four equations explicitly):
\begin{subequations}
\begin{eqnarray}
	&& 
	(1\!-\!x) \frac{d}{dx} 
	\phi_1^{(r)}
	= - \frac{1}{x} \frac{H(x,y)}{h_1(x)} \phi_3^{(r)}
	\\ && \hspace{5mm}
	+
	\left[
	(a_1 \!+\! b_1) \!-\! \frac{c}{x}
	\right] \phi_1^{(r-1)}
	+ 
	\left[ 
	\frac{a_1 r_1+b_1 p_1}{q} 
	\right]  
	\frac{1}{h_1(x)} \omega_0^{(r-1)}
	\!+\! 
	\frac{1}{h_1(x)} a_1 b_1 \omega_0^{(r-2)} 
	\;,
	\label{f3:3a}
	\nonumber \\ [1cm]
	&& 
	(1\!-\!y) \frac{d}{dy} 
	\phi_2^{(r)}
	= - \frac{1}{y} \frac{H(x,y)}{h_2(y)} \phi_3^{(r)}
	\\ && \hspace{5mm}
	+
	\left[
	(a_2 \!+\! b_2) \!-\! \frac{c}{y}
	\right] \phi_2^{(r-1)}
	+ 
	\left[ 
	\frac{a_2 r_2+b_2 p_2}{q} 
	\right]  
	\frac{1}{h_2(y)} \omega_0^{(r-1)}
	\!+\! 
	\frac{1}{h_2(y)} a_2 b_2 \omega_0^{(r-2)} 
	\;,
	\nonumber 
	\label{f3:3b}
	\end{eqnarray}
	\label{F3:transformed1}
\end{subequations}
\begin{subequations}
	\begin{eqnarray}
	&& 
	(xy\!-\!x\!-\!y) \frac{d}{dx} 
	\phi_3^{(r)}  
	=
	\left[
	(1\!-\!y)(a_1 \!+\! b_1) \!-\! \frac{y}{x} (a_2 \!+\! b_2  \!-\! c)
	\right] \phi_3^{(r-1)}
	\nonumber \\ && \hspace{5mm}
	+ \frac{a_1 r_1 + b_1 p_1}{q} (1\!-\!y) \frac{h_2(y)}{H(x,y)} \phi_2^{(r-1)}  
	+ a_1 b_1 (1\!-\!y) \frac{h_2(y)}{H(x,y)}  \phi_2^{(r-2)}
	\nonumber \\ && \hspace{5mm}
	- \frac{a_2 r_2 + b_2 p_2}{q}  \frac{y}{x} \frac{h_1(x)}{H(x,y)}  \phi_1^{(r-1)} 
	- a_2 b_2 \frac{y}{x}  \frac{h_1(x)}{H(x,y)}  \phi_1^{(r-2)} 
	\;,
	\label{f3:4a}
	\\ [1cm]
	&& 
	(xy\!-\!x\!-\!y) \frac{d}{dy} 
	\phi_3^{(r)}  
	=
	\left[
	(1\!-\!x)(a_2 \!+\! b_2) \!-\! \frac{x}{y} (a_1 \!+\! b_1  \!-\! c)
	\right] \phi_3^{(r-1)}
	\nonumber \\ && \hspace{5mm}
	+ \frac{a_2 r_2 + b_2 p_2}{q} (1\!-\!x) \frac{h_1(x)}{H(x,y)} \phi_1^{(r-1)}  
	+ a_2 b_2 (1\!-\!x) \frac{h_1(x)}{H(x,y)}  \phi_1^{(r-2)}
	\nonumber \\ && \hspace{5mm}
	- \frac{a_1 r_1 + b_1 p_1}{q}  \frac{x}{y} \frac{h_2(y)}{H(x,y)}  \phi_2^{(r-1)} 
	- a_1 b_1 \frac{x}{y}  \frac{h_2(y)}{H(x,y)}  \phi_2^{(r-2)} 
	\;,
	\label{f3:4b}
	\end{eqnarray}
	\label{F3:transformed2}
\end{subequations}
This is a system of linear differential equations 
with algebraic coefficients. For completeness, 
it should be supplemented by the boundary condition. 
Boundary conditions are discussed below. 

\noindent 
{\bf Remark}:\\
Let us recall that the surface of singularities $L(x,y)$ of the $F_3$ hypergeometric function is defined by the system of equations 
Eq.\ (\ref{F3:system}) and has the form
\begin{equation}
L: L_1 \cup L_2 \cup L_3 \cup L_4 \cup L_5
   \equiv x \cup y \cup (1-x) \cup (1-y) \cup (x+y-xy) \;.
\label{L}
\end{equation}
The extra functions, Eqs.(\ref{h}) and (\ref{H}), can be 
understood as  the ratio of  elements $L_a$ of Eq.\ (\ref{L}), and $q$-root is related with 
angle of rotations of curves $L_j$ around zero.

Indeed, a multiple polylogarithm can be understood as a
smooth map $U$  from one region of singularities, where the solution of differential equation exists, to another one:  
$\Li{A}{L_i} \xrightarrow{U} \Li{A}{L_j}$. Such map is nothing but an analytic continuation that mixed the singularities of the differential system Eq.\ (\ref{F3:system}). 

\subsubsection{Boundary conditions}
\label{F3:boundary}
The boundary conditions for the system of equations, 
Eqs.\ (\ref{F3:transformed1}),(\ref{F3:transformed2}), are defined by the  series representation, so that  
\begin{eqnarray} 
\left. 
\omega_0(z_1,z_2) \right|_{z_1=0} 
= {}_2F_1(a_2,b_2;c;z_2) \;, 
\quad 
\left. 
\omega_0(z_1,z_2) \right|_{z_2=0} 
=  {}_2F_1(a_1,b_1;c;z_1) \;. 
\end{eqnarray}
Keeping in mind that 
\begin{eqnarray}
\omega_j(z_1,z_2) & = &  
z_j \frac{a_j b_j}{c} 
\left. 
F_3(a_1, a_2, b_1, b_2, 1+c; z_1,z_2)\right|_{\substack{
	a_j \to a_j+1 \\
    b_j \to b_j+1}} \;, \quad j = 1,2 \;, 
\nonumber \\ 
\omega_3(z_1,z_2) 
& = & 
z_1 z_2 \frac{a_1 a_2 b_1 b_2}{c(1+c)} 
F_3(1+a_1, 1+a_2, 1+b_1, 1+b_2; 2+c; z_1,z_2) \;,
\nonumber 
\end{eqnarray}
we have
\begin{eqnarray}
\left.
\omega_1(z_1,z_2) 
\right|_{z_1=0}
& = & 0 \;, 
\quad 
\left. 
\omega_1(z_1,z_2) 
\right|_{z_2=0}
=  z_1 \frac{a_1 b_1}{c} {}_2F_1(1+a_1,1+b_1;1+c;z_1) \;, 
\nonumber \\ 
\left.
\omega_2(z_1,z_2) 
\right|_{z_2=0}
& = & 0 \;, 
\quad 
\left. 
\omega_2(z_1,z_2) 
\right|_{z_1=0}
=  z_2 \frac{a_2 b_2}{c} {}_2F_1(1+a_2,1+b_2;1+c;z_2) \;, 
\nonumber \\ 
\left. 
\omega_3(z_1,z_2) 
\right|_{z_1=0}
& = & 
\left. 
\omega_3(z_1,z_2) 
\right|_{z_2=0}
= 0 \;. 
\end{eqnarray}
The construction of the all-order $\varepsilon$-expansion of Gauss hypergeometric functions around rational values of parameters in terms of 
multiple polylogarithms 
has been constructed in Refs.\ \cite{kwy2006},\cite{kk2008}.

\subsubsection{The rational parametrization: 
	            towards multiple polylogarithms}
It is well-known that one-fold iterated integrals over 
algebraic functions are not, in general, expressible 
in terms of multiple polylogarithms 
but demand the introduction 
of a new 
class of functions~\cite{non-polylog:1,non-polylog:2,non-polylog:3}
\footnote{Originally, 
	such types of functions have been introduced in 
Ref.\ \cite{Bonciani}.}.

The iterative solution of the system Eq.~(\ref{F3:transformed1}),(\ref{F3:transformed2}), 
have the following form. 
In the first two orders of $\varepsilon$-expansion, we have
(these results follow from the series representation and a 
special choice of parameters):
\begin{subequations}
	\begin{eqnarray}
	&& 
	\omega_0^{(0)} = 1 \;, 
	\quad 
	\phi_1^{(0)} = \phi_2^{(0)} = \phi_3^{(0)} = 0 \;, 
	\\ && 
	\omega_0^{(1)} =  \phi_1^{(1)} = \phi_2^{(1)} = \phi_3^{(1)} = 0 \;, 
	\end{eqnarray}
\end{subequations}

The second iteration produces: 
\begin{subequations}
\begin{eqnarray}
\phi_3^{(2)}(x,y) & = & 0 \;, 
\\  
\phi_1^{(2)}(x,y) & = & a_1 b_1 \int_0^x \frac{dt}{(1-t) h_1(t)}  
 \equiv a_1 b_1 R_1(x) \;, 
\\  
\phi_2^{(2)}(x,y) & = & a_2 b_2 \int_0^y \frac{dt}{(1-t) h_2(t)}  
 \equiv  a_2 b_2 R_2(y)
\;, 
\\  
\omega_0^{(2)}(x,y) & = &  
\int_0^x \frac{dt_1}{t_1} \phi_1^{(2)} + {}_2F_1^{(2)}(x)
+ 
\int_0^y \frac{dt_1}{t_1} \phi_2^{(2)} + {}_2F_1^{(2)}(y)
\\ && 
=  
\left[ a_1 b_1 \int_0^x \frac{dt_1}{d_1} R_1(t) + a_2 b_2 \int_0^y \frac{dt_1}{t_1} R_2(t) \right]  
+ {}_2F_1^{(2)}(x) + {}_2F_1^{(2)}(y)
\nonumber 
\;, 
\end{eqnarray}
\end{subequations}
where ${}_2F_1^{(2)}(t)$ are the functions coming from boundary conditions.  
The finite part of the $F_3$ function is (in terms of $R$-functions):

	\begin{eqnarray}
	\phi_3^{(3)}(x,y) & = &  
	-\left( \frac{a_1 r_1 + b_1 p_1}{q}  \right)  a_2 b_2 
	h_2(y) R_2(y) \int_0^x \frac{dt}{H(t,y)\left(t+\frac{y}{1-y} \right)} 
	\nonumber \\ && 
	+ \left( \frac{a_2 r_2 + b_2 p_2}{q} \right)  a_1 b_1 
	\int_0^x dt \frac{h_1(t) R_1(t)}{H(t,y)}
	\left(
	\frac{1}{t} - \frac{1}{t+\frac{y}{1-y}}
	\right)
	\nonumber \\ && 
	- \left( \frac{a_2 r_2 + b_2 p_2}{q}  \right) a_1 b_1 h_1(x) R_1(x)
	\int_0^y dt \frac{dt}{H(x,t)\left(t+\frac{x}{1-x} \right)} 
	\nonumber \\ && 
	+ \left( \frac{a_1 r_1 + b_1 p_1}{q} \right) a_1 b_1 
	\int_0^y dt \frac{h_2(t) R_2(t)}{H(x,t)}
	\left(
	\frac{1}{t} - \frac{1}{t+\frac{x}{1-x}}
	\right)
	\;. 
	\end{eqnarray}

Up to some factor, the function
$R(t)$ coincides with the Gauss hypergeometric function with a rational set of parameters: 
$$
R(z)
\equiv
\int_0^z \frac{dt}{(1-t)} \frac{1}{h(t)}
\sim
{}_2F_1
\left(
\frac{r_j}{q}, \frac{p_j}{q},
1-\frac{p}{q}; z
\right)\;, \quad j = 1,2,
$$
where $p_j r_j = 0$. 
The $\varepsilon$-expansion of the Gauss hypergeometric function
around rational values of parameters
has been analyzed in detail in Ref.\ \cite{kk2008}.
It was shown that only the following cases are relevant: 
\begin{itemize}
	\item
	Integer set:
	$	
	r_j = p_j = p = 0  \;, 
	$
	\item
	Zero-balance type: 
	$r_j + p_j = -p \;, $
	\quad 
	$r_j p_j = 0 \;.$ 
	\item
	Binomial type: 
	$p=p_j = 0 $\;, \quad  $r_j \neq 0 $ \;, 
	(and symmetric one: 	
	$p=r_j = 0 $\;, \quad $p_j \neq 0 $)
	\item
	Inverse binomial type:
	$r_j=p_j = 0 $\;, \quad $p \neq 0 $\;. 
	\item
	Full type: 
	$r_j=p_j = -p \;.$
\end{itemize}
For each particular set of parameters, 
the rational parametrization of Eqs.~(\ref{F3:transformed1}),(\ref{F3:transformed2})
should be cross-checked. 
For illustration,  let us consider 
a few examples. 
\subsubsection{The rational parametrization: set 1}
\label{F3:set1}
Let us consider the hypergeometric function
$$
F_3\left(a_1 \varepsilon, 
         a_2  \varepsilon, 
         b_1\varepsilon, 
         b_2\varepsilon, 
         1-\frac{p}{q}+c\varepsilon; x,y \right) \;, 
$$
so that 
$$
r_1 = p_1 = p_2 = s_2 = 0 \;, \quad p \neq 0 \;. 
$$
For this set of parameters, we have 
$
s_1  =  0 \;, s_2 = 0 \;,  p \neq 0 \;, 
$
and 
$$
h(x) = \left( \frac{x}{x-1} \right)^{\frac{p}{q}} \;, 
\quad 
h(y)  =  \left( \frac{y}{y-1} \right)^{\frac{p}{q}} \;, 
\quad 
H(x,y)  =   
\left[ 
\frac{xy}
{((x-1)(y-1)-1)}
\right]^\frac{p}{q} \;.
$$
There is a parametrization 
that converts the  functions $h_1$ and $h_2$ 
into rational functions $P_1$ and $P_2$:
$$
\frac{x}{x-1} = P_1^q(\xi_1,\xi_2) \;, 
\quad 
\frac{y}{y-1} = P_2^q(\xi_1,\xi_2) \;, 
$$
where the functions $P_1$ and $P_2$ have the form 
\begin{equation}
P_m(x,y) = \Pi_{i,j,k,l} \frac{(x-a_i)(y-b_j)}{(x-c_k) (y-d_l)}\;, \quad 
m=1,2 \;, 
\label{rational}
\end{equation}
for a set of algebraic numbers $\{a_i,b_j,c_k,d_l\}$. 
In terms of new variables, the function $H$ has the form 
$$
H(x,y) = \frac{xy}{x+y-xy} = \frac{1}{\frac{1}{x} + \frac{1}{y} - 1} = 
\frac{1}{\frac{1}{P_1^q} + \frac{1}{P_2^q} + 1} = P_3^q \;,
$$
where 
$P_3$ 
is again a rational function of two variables of the type (\ref{rational}).

After a redefinition 
$
\frac{1}{P_i^q}  = Q_i^q \;, \quad i =1,2,3
$
we obtain the result  that a 
statement about existence of a rational parametrization for the functions 
$h(x),~h(y)$ and $H(x,y)$ is equivalent to the existence of 
three rational functions of two 
variables satisfying the equation  
\begin{eqnarray}
Q_1^q + Q_2^q + Q_3^q = 1\;.
\label{rational-parametrization}
\end{eqnarray}
To our knowledge, for $q>2$, a solution exists only in terms 
of elliptic functions. However, it may happen that 
such a parametrization exists for $q=2$, but we are not able to find it.

This problem is closely related to  
the solution of a functional equation. 
For example, for the equation 
$$
f^n + g^n = 1\;, 
$$
the solution can be characterized as follows: 
\cite{gross,baker}:
\begin{itemize}
\item
For $n=2$, all solutions are the form 
$$
f = \frac{2 \beta(z)}{ 1+\beta^2(z)} \;, 
\quad 
g = \frac{1-\beta^2(z)}{1+\beta^2(z)} \;,
$$
where $\beta(z)$ is an arbitrary function. 
\item
For $n=3$, one solution is given by
\begin{equation}
f = \frac{1}{2 \wp}
\left(
1 + \frac{1}{\sqrt{3}} \wp'
\right)\;, 
\quad 
g =  - \frac{1}{2 \wp}
\left(
1 - \frac{1}{\sqrt{3}} \wp'
\right)\;, 
\end{equation}
where $\wp$ is the Weierstrass $\wp$-function satisfy 
$
\left( \wp' \right)^2 = 4 \wp^3 - 1 \;. 
$
For $n=3$, the original equation is of genus $1$, so that 
uniformization theorem assures the existence of an elliptic solution. 
\end{itemize}

One of the most natural sets of variables for
the set of parameters under consideration is the following: 
$
P_1 = \xi_1 \;, P_2 = \xi_2 \;, 
$
so that 
\begin{equation}
H = \frac{1}{\xi_1^p \xi_2^p} 
\left(\xi_1^q \xi_2^q - \xi_1^q - \xi_2^q \right)^\frac{p}{q} \;. 
\label{surface}
\end{equation}

\subsubsection{The rational parametrization: set 2}
\label{F3:set2}
In a similar manner, let us analyze another set of parameters: 
$$
F_3\left(-\frac{p}{q} + a_1 \varepsilon, 
         a_2 \varepsilon, 
         b_1 \varepsilon, 
         -\frac{p}{q}+b_2\varepsilon, 
         1-\frac{p}{q}+c\varepsilon; x,y
\right) \;.
$$
In this case, 
$
s_1 = s_2 = -p \;,   
$
and the functions $h$ have the form
$
h(x) = x^\frac{p}{q} \;, 
\quad 
h(y) = y^\frac{p}{q} \;. 
$
Applying the same trick with the introduction of new 
functions $P_1$ and $P_2$, we would find that the existence of a rational parametrization corresponds in the present case to the 
validity of Eq.\ (\ref{rational-parametrization}). 
In particular, by introducing new variables 
$
x^\frac{1}{q} = \xi_1, y^\frac{1}{q} = \xi_2 \;, 
$ 
we obtain
$
H = 
\left(\xi_1^q \xi_2^q - \xi_1^q - \xi_2^q \right)^\frac{p}{q} \;. 
$
\subsubsection{The rational parametrization: set 3}
\label{F3:set3}
Let us analyze the following set of parameters: 
$
p = 0 \;,  s_1, s_2 \neq 0 \;, 
$
corresponding to 
$$
F_3\left( \frac{p_1}{q} + a_1 \varepsilon, 
          \frac{p_2}{q} + b_1 \varepsilon, 
                 b_2 \varepsilon, 
                 1+c\varepsilon; x,y
\right) \;.
$$
In this case, 
$$
h_1(x) = \left(1-x \right)^\frac{p_1}{q} \;, 
\quad 
h_2(y) = \left(1-y \right)^\frac{p_2}{q} \;, 
\quad 
H(x,y) = \left(
\frac{x^{s_2} y^{s_1}}{(xy-x-y)^{s_1+s_2}} 
\right)^\frac{1}{q} \;. 
$$
For simplicity, we set $s_1=-s_2=s$, and put 
$1-x = P_1^q$ and $1-y= P_2^q$, so that 
$H = \left(\frac{1-P_2^q}{1-P_1^q} \right)^{\frac{s}{q}} \equiv P_3$. In particular, for $P_1 = \xi_1, P_2 = \xi_2$, 
$H = \left(\frac{1-\xi_2^q}{1-\xi_1^q} \right)^{\frac{s}{q}}$. 

\subsubsection{The rational parametrization: set 4}
\label{F3:set4}
Let us analyze the following set of parameters: 
$
s_1 = -p \;, s_2 = 0 \;, 
$
so that hypergeometric function is 
$$
F_3\left(-\frac{p_1}{q} + a_1 \varepsilon, 
                          a_2 \varepsilon, 
b_1 \varepsilon, 
b_2 \varepsilon, 1-\frac{p}{q}+c\varepsilon; x,y
\right) \;.
$$
In this case, 
\begin{eqnarray}
h_1(x) & = & x^\frac{p}{q} \;, 
\quad 
h_2(y)  =  \left(\frac{y}{y-1} \right)^\frac{p}{q} \;, 
\quad 
H(x,y)  =   x^\frac{p}{q} \equiv h(x) \;.
\end{eqnarray}
Let us take a new set of variables $(x,y) \to (\xi_1,\xi_2)$: 
\begin{eqnarray}
\xi_1 & = & x^\frac{1}{q} \;, 
\quad 
\xi_2  =  \left(\frac{y}{y-1}\right)^\frac{1}{q} \;, 
\end{eqnarray}
In terms of a new variables we have: 
\begin{eqnarray}
H(x,y) & \equiv & h_1(x) =  \xi_1^p \;, 
\quad 
h_2(y)  =  \xi_2^p \;, 
\quad 
x+y-xy = 
\frac{\xi_1^q-\xi_2^q}{1-\xi_2^q} \;. 
\end{eqnarray}
Thus, a rational parametrization exists. 

\subsubsection{The rational parametrization: set 5}
\label{F3:set5}
Consider the set of parameters defined by 
$
s_2 = p = 0 \;,  
s_1 \neq 0,
$
that corresponds to the case
$
F_3\left(-\frac{p_1}{q} + a_1 \varepsilon, 
                          a_2 \varepsilon, 
b_1 \varepsilon, 
b_2 \varepsilon, 
1+c \varepsilon; x,y
\right) \;.
$
In this case, 
\begin{eqnarray}
h_1(x) = \left(1-x\right)^{\frac{p_1}{q}} \;, 
\quad 
h_2(y) = 1\;, 
\quad 
H(x,y) = \left( \frac{y}{xy-x-y} \right)^\frac{p_1}{q} \;, 
\end{eqnarray}
Let us suggest that 
$$
1-x = P^q \;, 
\quad 
H(x,y) = \frac{y}{x+y-xy} = Q^q \;,
$$
where $P$ and $Q$ are rational functions. 
Then, 
$
y =  \frac{(1-P^q)Q^q}{1-P^q Q^q} \;.
$
After the redefinition,
$
P Q = R \; 
$,  we get
$ 
y =  \frac{(1-P^q)}{(1-R^q)} \frac{R^q}{P^q} \;. 
$
The simplest version of $P$ and $R$ are polynomials: 
$
P = \xi_1 \;, 
$
and 
$ 
R = \xi_2 \;. 
$ 
In this parametrization, 
$$
h(x) = \xi_1^{p_1} \;, 
\quad 
H(x,y) = \left( \frac{\xi_2}{\xi_1} \right)^{p_1}\;, 
\quad 
y = \frac{1-\xi_1^q}{1-\xi_2^q} \left( \frac{\xi_2}{\xi_1}\right)^q  \;. 
$$
In this case, the rational parametrization exists. 

\subsubsection{Explicit construction of expansion: integer values of parameters}
Let us consider the construction of the $\ep$-expansion around 
integer values of parameters. 
If we put 
$$
\omega_0 = F_3(a_1 \ep, b_1 \ep, a_2 \ep, b_2 \ep, 1+c\ep;x,y), 
$$
then the system of differential equations can be presented 
in the form 
\begin{eqnarray}
\frac{\partial}{\partial x} \omega_1 & = & 
\left[ 
\frac{1}{x-1}
- \frac{1}{x} 
\right]
\omega_3 
- \left[ \frac{c}{x} + \frac{(a_1+b_1-c)}{x-1} \right] \ep \omega_1 
- a_1 b_1 \frac{1}{x-1} \ep^2 \omega_0 
\;, 
\\ 
\frac{\partial}{\partial y} \omega_2 & = & 
\left[ 
\frac{1}{y-1}
- \frac{1}{y} 
\right] \omega_3 
- \left[ \frac{c}{y} + \frac{(a_2+b_2-c)}{y-1} \right] \ep \omega_2 
- a_2 b_2 \frac{1}{y-1} \ep^2 \omega_0 
\;, 
\\
\frac{\partial}{\partial x} \omega_3 
& = &  
\left[ 
\frac{(a_2\!+\!b_2\!-\!c) }{x}
\!-\! 
\frac{(a_1 \!+\! b_1 \!+\! a_2\!+\!b_2\!-\!c)}{x+\frac{y}{1-y}} 
\right] \ep \omega_3 
\nonumber \\ && 
- 
\frac{a_1 b_1}{x+\frac{y}{1-y}} \ep^2 \omega_2 
\!+\! 
\left[ 
\frac{1}{x} 
- 
\frac{1}{x+\frac{y}{1-y}}
\right]
a_2 b_2 \ep^2 \omega_1 
\;, 
\\ 
\frac{\partial}{\partial y} \omega_3 
& = &  
\left[ 
\frac{(a_1\!+\!b_1\!-\!c) }{y}
\!-\! 
\frac{(a_1 \!+\! b_1 \!+\! a_2\!+\!b_2\!-\!c)}{y+\frac{x}{1-x}} 
\right] \ep \omega_3 
\nonumber \\ && 
- 
\frac{a_2 b_2}{y+\frac{x}{1-x}} \ep^2 \omega_1 
\!+\! 
\left[ 
\frac{1}{y} 
- 
\frac{1}{y+\frac{x}{1-x}}
\right]
a_1 b_1 \ep^2 \omega_2 
\;. 
\end{eqnarray}
This system can be straightforwardly integrated in terms of multiple polylogarithms, defined via a one-fold 
iterated integral $G$, where 
\begin{eqnarray}
G(z;a_k,\vec{a}) & = & 
\int_0^{z} \frac{dt}{t-a_k}
G(t;\vec{a}) \;.
\label{G}
\end{eqnarray}
 
In addition, the $\ep$-expansion of a Gauss hypergeometric function around integer values of parameters is needed. 
It has the following form (see Eq.~(34) in ~\cite{MKL:Gauss}):
\begin{eqnarray}
&& 
{}_2F_1(a\ep,b\ep;1\!+\!c\ep;z) 
= 
1  \!+\! 
a b \ep^2 \Li{2}{z}
\nonumber \\ && \hspace{5mm}
+ 
a b \ep^3 
\left[(a \!+\! b \!-\! c) \Snp{1,2}{z} \!-\! c \Li{3}{z} \right] 
+ O(\ep^4)
\;.
\nonumber \\ 
\end{eqnarray}

The first iteration gives rise to
\begin{eqnarray}
\omega_0^{(0)} & = & 1 \;, 
\quad 
\omega_1^{(0)} = \omega_2^{(0)} = \omega_3^{(0)} = 0 \;, 
\quad 
\omega_0^{(1)} = \omega_1^{(1)}  =  \omega_2^{(1)} = \omega_3^{(1)} = 0 \;. 
\nonumber 
\end{eqnarray}

The results of the second iteration are the following: 
\begin{eqnarray}
\omega_3^{(2)} & = & 0 \;, 
\quad 
\omega_1^{(2)}  =  -a_1 b_1 \ln(1-x) \;, 
\quad 
\omega_2^{(2)}  = - a_2 b_2 \ln(1-y) \;, 
\nonumber \\ 
\omega_0^{(2)} & = &  a_1 b_1 \Li{2}{x} + a_2 b_2 \Li{2}{y} \;, 
\end{eqnarray}
where the classical polylogarithms $\Li{n}{z}$ 
are defined as 
\begin{equation}
\Li{1}{z} = - \ln(1-z) \; ,
\quad 
\Li{n+1}{z} = \int_0^z \frac{dt}{t} \Li{n}{t}, \quad n\ge 1.
\label{polylogarithm}
\end{equation}

After the third iteration, we have
\begin{eqnarray}
\omega_{3}^{(3)} &  = & 0
\\ 
\omega_1^{(3)} & = &  
\frac{1}{2} a_1 b_1 (a_1+b_1-c) \ln^2 (1-x) 
- a_1 b_1 c \Li{2}{x} 
\;, 
\\ 
\omega_2^{(3)} & = &  
\frac{1}{2} a_2 b_2 (a_2+b_2-c) \ln^2 (1-y) 
- a_2 b_2 c \Li{2}{y} 
\;, 
\\
\omega_0^{(3)} & = &  
- a_1 b_1 c \Li{3}{x} 
- a_2 b_2 c \Li{3}{y} 
\nonumber \\ && 
+ a_1 b_1 (a_1+b_1-c) \Snp{1,2}{x}
+ a_2 b_2 (a_2+b_2-c) \Snp{1,2}{y}
\;,
\end{eqnarray}
where $\Snp{a,b}{z}$ are the Nielsen polylogarithms: 
\begin{equation}
z \frac{d}{d z} \Snp{a,b}{z}  =  \Snp{a-1,b}{z} \;, 
\quad 
\Snp{1,b}{z}  =  \frac{(-1)^b}{b!} \int_0^1 \frac{dx}{x} \ln^b(1-zx) \;. 
\label{nielsen}
\end{equation}

The result of the next iteration, $\omega_3^{(4)}(x,y)$, 
can be expressed in several equivalent forms: 
\begin{eqnarray}
\frac{\omega_3^{(4)}(x,y)}{a_1 a_2 b_1 b_2} 
& = &  
\ln(1-y) G_1\left(x; -\frac{y}{1-y} \right)
- G_2(x;1)
+ G_{1,1}\left(x; - \frac{y}{1-y},1 \right) 
\nonumber \;, 
\\ 
\frac{\omega_3^{(4)}(x,y)}{a_1 a_2 b_1 b_2 } 
& = & 
\ln(1-x) G_1\left(y; -\frac{x}{1-x} \right)
- G_2(y;1)
+ G_{1,1}\left(y; - \frac{x}{1-x},1 \right)
\;. 
\nonumber 
\end{eqnarray}
Keeping in mind that 
\begin{eqnarray}
&& 
G_{1,1}\left(x; - \frac{y}{1-y},1 \right)
= 
\int_0^x \frac{dt}{t+\frac{y}{1-y}} \ln (1-t) 
\\ && 
=  
- \ln (1-y) \ln (x+y-xy) 
+ \ln(1-y) \ln y
- \Li{2}{x+y-xy} 
+ \Li{2}{y} 
\nonumber  \;,
\end{eqnarray}
the result can be written in a very simple form,
\begin{eqnarray}
\frac{\omega_3^{(4)}(x,y)}{a_1 a_2 b_1 b_2} 
= \Li{2}{x} + \Li{2}{y} - \Li{2}{x+y-xy} \;.  
\end{eqnarray}

Taking into account that 
$
\frac{\omega_3^{(4)}(x,y)}{a_1 a_2 b_1 b_2}
=  \frac{1}{2} xy F_3(1,1,1,1;3;x,y)\;, 
$
we obtain the well-known result~\cite{sanchis-lozano} 
$$
\frac{1}{2} xy F_3(1,1,1,1;3;x,y) = \Li{2}{x} + \Li{2}{y} - \Li{2}{x+y-xy} \;.
$$
There is also another form ~\cite{brychkov} for this integral,
\begin{eqnarray}
\frac{1}{2} xy F_3(1,1,1,1;3;x,y) &  = & 
\Li{2}{\frac{x}{x+y-xy}} 
- 
\Li{2}{\frac{x-xy}{x+y-xy}} 
\nonumber \\ && 
- \ln (1-y) \ln \left(\frac{y}{x+y-xy} \right) \;.
\nonumber 
\end{eqnarray}
One form can be converted to the other using the well-known 
dilogarithm identity 
$$
\Li{2}{\frac{1}{z}}
= 
-\Li{2}{z} 
- \frac{1}{2} \ln^2 (-z)
-\zeta_2 \;,
$$
together with the attendant functional relations~\cite{lewin,Li2identities}. 

The following expressions result from direct iterations in terms of $G$-functions: 
\begin{eqnarray}
&&
\frac{\omega_1^{(4)}(x,y)}{a_1 b_1} = 
a_2 b_2 
\Biggl\{ 
\ln(1-y) 
\left[
G_{1,1}\left(x; 1, -\frac{y}{1-y} \right)
- 
G_{2}\left(x; -\frac{y}{1-y} \right)
\right]
\nonumber \\ && 
- G_{1,2}(x;1,1) + G_{3}(x;1)
+ G_{1,1,1}\left(x; 1, - \frac{y}{1-y},1 \right)  
- G_{2,1}\left(x; - \frac{y}{1-y},1 \right) 
\Biggr\} 
\nonumber \\ && 
+ a_2 b_2 G_1(x;1) G_2(y;1) 
- c \Delta_1 G_{2,1}(x;1,1)
- \Delta_1^2 G_{1,1,1}(x;1,1,1)
- c^2 G_{3}(x;1)
\nonumber \\ && 
+ (a_1 b_1 \!-\! c \Delta_1) G_{1,2}(x;1,1) 
\;, 
\end{eqnarray}
where 
$$
\Delta_j  = a_j + b_j - c \;.
$$
The $G$-functions can be converted into classical or 
Nielsen polylogarithms with the help of the following relations: 
\begin{subequations}
\begin{eqnarray}
&& 
G_{1,1}
\left( 
x; 1, - \frac{y}{1-y}
\right)
= 
G_1(x; 1)
G_1\left( 
x; - \frac{y}{1-y}
\right)
- 
G_{1,1}\left( 
x; - \frac{y}{1-y}, 1
\right) \;,
\\ && 
G_{1}
\left( 
x; 1
\right)
G_{1,1}
\left( 
x; - \frac{y}{1-y}, 1
\right)
= 
G_{1,1,1}
\left( 
x; 1, - \frac{y}{1-y}, 1
\right)
\\ && \hspace{40mm}
+ 
2G_{1,1,1}
\left( 
x; - \frac{y}{1-y}, 1, 1
\right) 
\;, 
\nonumber \\ && 
G_{1,1,1}
\left( 
x; - \frac{y}{1-y}, 1, 1
\right) 
= 
\Snp{1,2}{x+y-xy}- \Snp{1,2}{y}
\\ && 
+ \frac{1}{2} \ln^2 (1-y)
\left[ 
\ln(x+y-xy) 
\!-\! 
\ln y 
\right]
+ \ln(1-y) 
\left[ 
\Li{2}{x+y-xy} \!-\! \Li{2}{y} 
\right]
\;, 
\nonumber \\ &&
G_{2,1}\left(x; -\frac{y}{1-y} ,1 \right)
+ \ln(1-y)  G_{2}\left(x; -\frac{y}{1-y} \right)
= 
G_{1,2}\left(1; 1-\frac{1}{y}, \frac{1}{x} \right)
+ G_{3}\left(x; 1 \right)
\nonumber \\  && 
= \int_0^x \frac{du}{u}
\left[ 
\Li{2}{y} - \Li{2}{u+y-uy} 
\right] \;.
\end{eqnarray}
\end{subequations}
In a similar manner, 
\begin{eqnarray}
&&
\frac{\omega_2^{(4)}(x,y)}{a_2 b_2} = 
a_1 b_1  
\Biggl\{ 
\ln(1-x) 
\left[
G_{1,1}\left(y; 1, -\frac{x}{1-x} \right)
- 
G_{2}\left(y; -\frac{x}{1-x} \right)
\right]
\nonumber \\ && 
- G_{1,2}(y;1,1) + G_{3}(y;1)
+ G_{1,1,1}\left(y; 1, - \frac{x}{1-x},1 \right)  
- G_{2,1}\left(y; - \frac{x}{1-x},1 \right) 
\Biggr\} 
\nonumber \\ && 
- c \Delta_2 G_{2,1}(y;1,1)
- \Delta_2^2 G_{1,1,1}(y;1,1,1)
- c^2 G_{3}(y;1)
- c \Delta_2 G_{1,2}(y;1,1)
\nonumber \\ && 
+ a_1 b_1 G_1(y;1) G_2(x;1) 
+ a_2 b_2 G_{1,2}(y;1,1) 
\;, 
\end{eqnarray}
and the last term, expressible in terms of functions of order $3$, is $\omega_3^{(5)}$. 

\subsubsection{Construction of $\ep$-expansion via 
	           integral representation}

An alternative approach to construction of the 
higher order $\ep$-expansion of generalized hypergeometric 
functions is based on their 
integral representation. 
We collect here some representations
for the Appell hypergeometric function $F_3$
extracted from Refs.\ \cite{slater} and ~\cite{srivastava}.

For our purposes, the most useful expression is the following:
(see Eq.\ (16) in Section 9.4. of ~\cite{srivastava}), 
~\cite{sanchis-lozano,brychkov}:
\begin{eqnarray}
&& 
\frac{\Gamma(c_1) \Gamma(c_2)}{\Gamma(c_1+c_2)}
F_3(a_1,b_1,a_2,b_2,c_1+c_2;x,y)
\nonumber \\ && 
 = \int_0^1 du u^{c_1-1} (1-u)^{c_2-1}
 {}_2F_1(a_1,b_1;c_1;ux)
 {}_2F_1(a_2,b_2;c_2;(1-u)y) \;.
 \label{F3:1}
\end{eqnarray}
Indeed, expanding one of the hypergeometric functions as a power series leads to 
$$
\int_0^1 u^{c_1-1} (1-u)^{j+c_2-1}
{}_2F_1(a_1,b_1;c_1;ux)
\sum_{j=0}^\infty
\frac{(a_2)_j (b_2)_j y^j}{(c_2)_j j!}
 \;. 
$$
The order of summation and integration can be interchanged 
in the domain of convergence of the series, and after integration we obtain Eq.\ (\ref{F3:1}).

The two-fold integral representation~\cite{slater},
\begin{eqnarray}
&& 
\frac{\Gamma(b_1) \Gamma(b_2)\Gamma(c-b_1-b_2)}{\Gamma(c)}
F_3(a_1,a_2;b_1,b_2;c;x,y)
= 
\\ && 
\int \int_{0 \leq  u, 0 \leq v, u+v \leq 1} du dv 
u^{b_1-1} v^{b_2-1}
(1-u-v)^{c-b_1-b_2-1}
(1-ux)^{-a_1}
(1-vy)^{-a_2} \;,
\nonumber 
\label{F3:slater} 
\end{eqnarray}
can be reduced to the following integral: 
\begin{eqnarray}
&& 
\frac{\Gamma(b_1) \Gamma(b_2)\Gamma(c-b_1-b_2)}{\Gamma(c)}
F_3(a_1,a_2;b_1,b_2;c;x,y)
\label{F3:2}
\\ && 
= 
\int_0^1 du \int_0^{1-u} dv 
u^{b_1-1} v^{b_2-1}
(1-u-v)^{c-b_1-b_2-1}
(1-ux)^{-a_1}
(1-vy)^{-a_2}
\nonumber \\ && 
= 
\frac{\Gamma(b_2) \Gamma(c-b_1-b_2)}{\Gamma(c-b_1)}
\nonumber \\ && \hspace{5mm}
\times 
\int_0^1 du u^{b_1-1} (1-ux)^{-a_1} (1-u)^{c-b_1-1}
~_2F_1
\left( \begin{array}{c|}
a_2, b_2 \\
c-b_1
\end{array}~ (1-u)y \right) \;.
\nonumber 
\end{eqnarray}

For the particular values of parameters $(c_1=a_1,c_2=a_2)$, the 
integral Eq.\ (\ref{F3:1}) can be reduced to the Appell function $F_1$: 
\begin{eqnarray}
F_3(a_1, a_2, b_1, b_2; a_1+a_2; x,y)
& = & 
\frac{1}{(1-y)^{b_2}}
F_1\left( a_1, b_1, b_2, a_1+a_2; x, -\frac{y}{1-y} \right) \;.
\nonumber \\ 
& = &  
\frac{1}{(1-x)^{b_1}}
F_1\left( a_2, b_1, b_2, a_1+a_2; -\frac{x}{1-x}, y  \right) \nonumber \;.
\end{eqnarray}

Using the one-fold integral representation for the Appell function $F_1$, 
it is possible to prove the following relations:  
\begin{eqnarray}
&& 
F_3(a,c-a,b,c-b,c,x,y)
= 
(1-y)^{c+a-b}
~_2F_1
\left( \begin{array}{c|}
a, b \\
c
\end{array}~ x+y-xy \right) \;.
\end{eqnarray}

Using Eqs.\ (\ref{F3:1}) and (\ref{F3:2}), the one-fold 
integral representation can be written for the coefficients of 
the $\ep$-expansion of the hypergeometric function $F_3$
via the $\ep$-expansion of the  Gauss hypergeometric function, constructed in Refs.\ \cite{kwy2006}, ~\cite{kk2008}.

The coefficients of the $\ep$-expansion of the Gauss hypergeometric function can be expressed in terms 
of multiple polylogarithms of a $q$-root of unity with arguments 
$\left(\frac{z}{z-1}\right)^\frac{1}{q}$,
$z^\frac{1}{q}$ or $(1-z)^\frac{1}{q}$ (see also ~\cite{nested2}), so that the problem of finding a rational parametrization reduces to the problem of finding a rational parametrization of the integral kernel of Eqs.\ (\ref{F3:1}) and (\ref{F3:2}) in terms of variables generated by the $\ep$-expansion of the Gauss hypergeometric function. 

The construction of the higher-order 
$\ep$-expansion of the Gauss hypergeometric function 
around rational values of parameters ~\cite{kwy2006,kk2008},
plays an crucial role in construction of the higher-order 
$\ep$-expansion of many (but not all)
Horn-type hypergeometric functions.

\subsubsection{Relationship to Feynman Diagrams}
Let us consider the one-loop pentagon with vanishing external
legs.
 The higher-order $\ep$-expansion for this diagram 
has been constructed~\cite{pentagon1} in terms of iterated one-fold integrals over algebraic functions. 
In Ref.\ \cite{pentagon2}, the hypergeometric representation 
for the one-loop 
pentagon with vanishing external momenta has been constructed as a sum of  
Appell hypergeometric functions $F_3$.

In ~\cite{BKK2012}, where our differential equation approach is presented, it was pointed out that 
the one-loop pentagon can be expressed in terms of multiple 
polylogarithms. Ref.\ \cite{pentagon4} verified the numerical agreement between the results of Refs.\ \cite{pentagon1} and 
\cite{pentagon2}, and Ref.\ \cite{pentagon5} constructed 
the iterative solution of the differential equation \cite{kotikov}.

Let us recall the results of Ref.\ \cite{pentagon2}.
The one-loop massless pentagon is expressible in terms of the Appell function $F_3$
with the following set of parameters: 
\begin{eqnarray}
\Phi_5^{(d)} \sim F_3\left( 1,1,\frac{7-d}{2},1,\frac{10-d}{2};x,y \right) \;, 
\label{pentagon1}
\end{eqnarray}
where $d$ is dimension of space-time. 
Another representation presented in 
\cite{pentagon2} has the structure
\begin{eqnarray}	
H_5^{(d)}
\sim 
F_3\left( \frac{1}{2},1,1,\frac{d-2}{2},\frac{d+1}{2};\frac{y}{x+y-xy}, \frac{1}{x} \right) 
\;.
\label{pentagon2}
\end{eqnarray}	

Let us consider the case of $d=4-2\ep$. 
The first representation, Eq.\ (\ref{pentagon1}), is 
$$
\Phi_5^{(4-2\ep)} \sim F_3\left( 1,1,\frac{3}{2}-\ep,1,4-\ep;x,y \right) \;. 
$$
This case, 
$$
\{
p_1 = p_2 =r_2 = p = 0 \}\;, 
\{r_1 = 1 \;, q=2 \}
\Longrightarrow
s_1 \neq 0; \quad s_2 = 0 \;, \quad p = 0 \;,
$$
corresponds to our {\bf set 5}, so that 
the {\bf $\ep$-expansion is expressible in terms of multiple  polylogarithms}, defined by Eq.~(\ref{G}). 

For the other representation, Eq.~(\ref{pentagon2}),   
$$	
H_5^{(4-2\ep)} 
\sim
	F_3\left( \frac{1}{2},1,1,\ep,\frac{5}{2}-\ep;\frac{y}{x+y-xy}, \frac{1}{x} \right) \;,  
$$
so that it is reducible to the following set of parameters: 
$$
\{p_2 = r_1 = r_2 = 0 \}\;, 
\quad 
\{p_1 = 1\;, q=2, \; p=1 \}\;
\Longrightarrow
	s_1 = 1; \quad s_2 = 0  \;, \quad p = -1 \;.
$$	
This is our {\bf set 4}, so that the
{\bf $\ep$-expansion is expressible in terms of multiple polylogarithms}, 
defined by Eq.~(\ref{G}).

The $\ep$-expansion of the one-loop pentagon
 about $d=3-2\ep$ could be treated in a similar manner. In this case, 
the first representation corresponds to {\bf set 1} 
$$
	\Phi^{(3-2\ep)} \sim  F_3\left( 1,1,2+\ep,1,\frac{7}{2}+\ep;x,y \right) 
\Longrightarrow
		p_1 = p_2 =r_1 = r_2 = 0, \quad 
			q=2 \;, p=1 
$$ 
and there is no rational parametrization, so that the result of the {\bf $\ep$-expansion is expressible in terms of a one-fold 
iterated integral over algebraic functions}.

The other representation, Eq.\ (\ref{pentagon2}), also 
cannot be expressed in terms of multiple polylogarithms:
$$
H_5^{(3-2\ep)} 	\sim   
	F_3\left( \frac{1}{2},1,1,\frac{1}{2} - \ep,2-\ep;\frac{y}{x+y-xy}, \frac{1}{x} \right) 
	\Longrightarrow
	\begin{array}{l}
	p_2 = r_1 = p = 0 \;, \\
	q=2 \; \quad p_1 = r_2 = 1\;, \\
	s_1 = 1; \quad s_2 = 1 \;.
	\end{array}
$$
This corresponds to {\bf set 3}, and the 
{\bf $\ep$-expansion is expressible in terms of one-fold iterated integral over algebraic functions}.

In this way, the question
of the all-order $\ep$-expansion of a one-loop Feynman diagram in terms of multiple polylogarithms is reduced to the question of the existence of 
a rational parametrization for the (ratio) of singularities.  

{\bf Remark}:
The dependence of the coefficients of the $\ep$-expansion (multiple polylogarithms or elliptic function)
on the dimension of space-time is not new.
In particular, it is well known that the two-loop sunset 
in $3-2\ep$ dimension is expressible in terms of polylogarithms~\cite{rajantie,lee}
and demands introduction of new functions in $4-2\ep$ dimension 
\cite{Laporta,tarasov:sunset,Bloch,Adams,Bloch:2016,Duhr,Bogner}.

\section{Conclusion}
The deep relationship between Feynman diagrams and hypergeometric functions has been reviewed, and we
have tried to enumerate all approaches and recent results on that subject. Special attention was devoted to the discussion of 
different algorithms for constructing the analytical 
coefficients of the $\ep$-expansion of multiple hypergeometric functions.
We have restricted ourselves to
multiple polylogarithms and functions related to 
integration over rational functions 
(the next step after multiple polylogarithms). 
The values of parameters related to elliptic polylogarithms 
was beyond our consideration. 

We have presented our technique for the construction of 
coefficients of the higher order $\ep$-expansion of multiple Horn-type hypergeometric functions, developed by the authors~\footnote{Unfortunately, the further 
	prolongation of this 
	project has not been supported by DFG, 
	so that many interesting results remain unpublished.} 
during the period 2006 -- 2013.
One of the main results of interest 
was the observation~\cite{DK1,DK2,DK3,kwy2006,kwy2007,kk2008,bkm2013} 
that for each Horn-type hypergeometric function, a 
set of parameters can be found so that the coefficients of the 
$\ep$-expansion include only functions of weight one
(so-called ``pure functions,''  in a modern terminology). 
As was understood in 2013 by Johannes Henn 
 ~\cite{Henn:1,Henn:2}, 
this property is valid not only for 
hypergeometric functions but also for generic Feynman diagrams. 

Our approach is based on the systematic analysis of the
system of hypergeometric differential equations (linear differential operators of hypergeometric type with polynomial coefficients) 
and does not demand the
existence of an integral representation, 
which is presently unknown for a large class
of multiple Horn-type hypergeometric functions (they could be deduced, 
but are not presently available in the mathematical literature). 

Our approach is based on the factorization of the system of 
differential equations into a product of differential operators,
together with finding a rational parametrization and constructing
iterative solutions.  
To construct such a system, an auxiliary manipulation with 
parameters (shifting by integer values) is required,
which can be done with the help of the HYPERDIRE set of 
programs~\cite{hyperdire}.
This technique
is applicable not only to 
hypergeometric functions defined by series but also to multiple 
Mellin-Barnes integrals~\cite{KK:MB} -- one of the representations of Feynman diagrams in a covariant gauge.
We expect that the present technique is directly applicable (with some 
technical modifications) to the construction of the $\ep$-expansion 
of hypergeometric functions 
beyond multiple polylogarithms, specially, that the two-loop sunset 
has a simple hypergeometric representation~\cite{tarasov:sunset}.   

There are two of our considerations that have not 
been solved algorithmically:
 (a) the factorization of linear partial differential operators into irreducible factors
is not unique, as has been ilustrated by Landau~\cite{landau} 
(see also Ref.\ \cite{schwarz}):
$
(\partial_x+1) (\partial_x+1) (\partial_x+x\partial_y)
= 
(\partial^2_x+x \partial_{xy}  
+ \partial_x+(2+x) \partial_y) (\partial_x+1) \; ; 
$
(b) The choice of parametrization is still an open problem, but there is essential progress in this direction~\cite{root1,root2}.

Our example has shown that such a parametrization is
defined by the locus of singularities of a system of differential 
equations, so that the problem of finding a rational parametrization 
is reduced to the parametrization of solutions of the
Diophantine equation for the singular locus of a Feynman diagram and/or hypergeometric function.
It is well known that in the case of a positive solution of this problem (which has no complete algorithmic solution), 
the corresponding system of partial differential equations 
of a few variables takes the simplest structure.
At the same time, there is a relationship 
between the type of solution of the Diophantine equation 
for the singular locus 
and the structure of the coefficients of the $\varepsilon$-expansion: 
a linear solution allows us to write the results of the
$\varepsilon$-expansion 
in terms of multiple polylogarithms. 
An algebraic solution gives rise to functions different from multiple polylogarithms and elliptic functions, {\em etc}. 
It is natural to expect that, in the case of an elliptic solution of the Diophantine equation
for the singular locus, the results for the $\varepsilon$-expansion are related to 
the elliptic generalization of multiple polylogarithms.

Another quite interesting and still algorithmically open problem is the transformation of multiple Horn-type hypergeometric functions with reducible monodromy to 
hypergeometric functions with irreducible monodromy. 
In the application to Feynman diagrams, such transformations correspond to functional relations, studied recently by Oleg Tarasov 
~\cite{Tarasov:functional1,Tarasov:functional2,Tarasov:functional3}, 
and by Andrei Davydychev~\cite{davydychev:box}.

Further analysis of the symmetries of the hypergeometric differential equations related to the Mellin-Barnes representation 
of Feynman diagrams (for simplicity we will call it the hypergeometric representation)  
has revealed their deep connection to the holonomic properties of Feynman diagrams. 
In particular, a simple and fast algorithm was constructed for the 
reduction of any Feynman diagram 
having a one-fold Mellin-Barnes integral representation to a set of master integrals~\cite{KK:MB}. 

The importance of considering the
dimension of the irreducible representation instead of generic 
holonomic rank has been pointed out in the application to Feynman 
diagrams~\cite{bkk2009}. In the framework of this approach, 
the set of irreducible, non-trivial master-integrals 
corresponds to the set of irreducible (with respect to analytical continuation of the variables, masses,
and external momenta) solutions of the corresponding system of hypergeometric differential equations, 
whereas diagrams expressible in terms of Gamma-functions correspond 
to Puiseux-type solutions (monomials with respect to Mandelstam variables) of the original system of hypergeometric equations.

\begin{acknowledgement}
    MYK is very indebted to Johannes Bl\"umlein, Carsten Schneider 
	and Peter Marquard for the
	invitation  and for creating such a stimulating atmosphere
	in the workshop ``Anti-Differentiation and the
	Calculation of Feynman Amplitudes'' at DESY Zeuthen.
	BFLW thanks Carsten Schneider for kind the hospitality at RISC.
    MYK thanks the Hamburg University for hospitality while 
	working on the manuscript. SAY acknowledges support from The Citadel Foundation.
\end{acknowledgement}

\end{document}